\documentclass[letterpaper,journal]{IEEEtran}
\usepackage[utf8]{inputenc}
\usepackage{amsmath,amsfonts}
\usepackage{algorithmic}
\usepackage{array}
\usepackage{textcomp}
\usepackage{stfloats}
\usepackage{url}
\usepackage{verbatim}
\usepackage{graphicx}

\usepackage{tabularx}
\usepackage{tabularray}
\usepackage{xcolor}
\usepackage{amssymb}

\usepackage{tikz}
\usepackage{amsmath}

\usepackage{balance}
\usepackage{url}

\usepackage{float}

\newcommand{\SIGL}[1]{\textsc{SIGL}}
\newcommand{\threatrace}[1]{\textsc{threaTrace}}
\newcommand{\nodlink}[1]{\textsc{NodLink}}
\newcommand{\magic}[1]{\textsc{Magic}}
\newcommand{\kairos}[1]{\textsc{Kairos}}
\newcommand{\flash}[1]{\textsc{Flash}}
\newcommand{\rcaid}[1]{\textsc{R-Caid}}

\newcommand{\fudd}[1]{\textsc{Fudd}}
\newcommand{\grasp}[1]{\textsc{Grasp}}
\newcommand{\orthrus}[1]{\textsc{Orthrus}}
\newcommand{\velox}[1]{\textsc{Velox}}
\newcommand{\darpa}[1]{\textsc{Darpa}}
\newcommand{\cadetsEthree}[1]{Cadets E3}
\newcommand{\cadets}[1]{Cadets}
\newcommand{\cadetsEfive}[1]{Cadets E5}
\newcommand{\theia}[1]{Theia}
\newcommand{\theiaEthree}[1]{Theia E3}
\newcommand{\theiaEfive}[1]{Theia E5}
\newcommand{\Clearscope}[1]{Clearscope}
\newcommand{\ClearscopeEthree}[1]{{Clearscope E3}}
\newcommand{\ClearscopeEfive}[1]{{Clearscope E5}}
\newcommand{\optc}[1]{{OpTC}}
\newcommand{\optcfirst}[1]{{OpTC 051}}
\newcommand{\optcsecond}[1]{{OpTC 201}}
\newcommand{\optcthrid}[1]{{OpTC 501}}

\newcommand{\meane}{\ensuremath{\mu_{e}}}
\newcommand{\meanh}{\ensuremath{\mu_{TP}}}
\newcommand{\hits}{\ensuremath{{TP}}}
\newcommand{\meana}{\ensuremath{\mu_{a}}}
\newcommand{\meanalarms}{\ensuremath{\mu_{a}}}

\newcommand{\cvalarms}{\ensuremath{\hat{c}_{\mathrm{V}a}}}
\newcommand{\cvADP}{\ensuremath{\hat{c}_{\mathrm{V}\mathrm{ADP}}}}
\newcommand{\attackrecall}[1]{\ensuremath{A_{recall}}}
\newcommand{\cvattackrecall}[1]{\ensuremath{\hat{c}_{\mathrm{V}\mathrm{A_{r}}}}}
\newcommand{\meanattackrecall}[1]{\ensuremath{\mu_{\mathrm{A_{r}}}}}

\newcommand{\gtgrasp}[1]{\ensuremath{GT_{\textsc{Grasp}}}}
\newcommand{\gtvelox}[1]{\ensuremath{GT_{\textsc{Velox}}}}
\newcommand{\gtE}[1]{\ensuremath{GT_{e}}}

\newcommand{\gt}[1]{GT}
\newcommand{\gts}[1]{GTs}

\newcommand{\cmark}{\checkmark}
\newcommand{\cstar}{\checkmark\textsuperscript{*}}

\definecolor{velox}{RGB}{0,0,0} 
\definecolor{cadetsfirst}{RGB}{127, 201, 127}
\definecolor{cadetssecond}{RGB}{190, 174, 212}
\definecolor{theiafirst}{RGB}{253, 192, 134}
\definecolor{theiasecond}{RGB}{0, 0, 0} 
\definecolor{clearscopefirst}{RGB}{56, 108, 176}
\definecolor{grasp}{RGB}{0,0,0} 
\definecolor{clearscopesecond}{RGB}{255, 204, 203} 
\definecolor{optcfirst}{RGB}{240, 2, 127}
\definecolor{optcsecond}{RGB}{191, 91, 22}
\definecolor{optcthird}{RGB}{102, 102, 102}
\definecolor{orthrus}{RGB}{0,0,0}

\newcommand{\cadetsfirstbox}[1]{%
  \raisebox{.7\height}{\colorbox{cadetsfirst}{#1}}%
}
\newcommand{\cadetssecondbox}[1]{%
  \raisebox{.7\height}{\colorbox{cadetssecond}{#1}}%
}
\newcommand{\theiafirstbox}[1]{%
  \raisebox{.7\height}{\colorbox{theiafirst}{#1}}%
}
\newcommand{\theiasecondbox}[1]{%
  \raisebox{.7\height}{\colorbox{theiasecond}{#1}}%
}
\newcommand{\clearscopefirstbox}[1]{%
  \raisebox{.7\height}{\colorbox{clearscopefirst}{#1}}%
}
\newcommand{\clearscopesecondbox}[1]{%
  \raisebox{.7\height}{\colorbox{clearscopesecond}{#1}}%
}
\newcommand{\optcfirstbox}[1]{%
  \raisebox{.7\height}{\colorbox{optcfirst}{#1}}%
}                       
\newcommand{\optcsecondbox}[1]{%
  \raisebox{.7\height}{\colorbox{optcsecond}{#1}}%
}
\newcommand{\optcthirdbox}[1]{%
  \raisebox{.7\height}{\colorbox{optcthird}{#1}}%
}

\newcommand{\Fone}[1]{\ensuremath{\mathrm{F}_1}}

\newcommand{\MacroFone}[1]{\ensuremath{\mathrm{M}\text{-}\mathrm{F}_1}}
\newcommand{\MacroFoneFull}[1]{\ensuremath{\mathrm{Macro}\text{-}\mathrm{F}_1}}

\newcommand{\meanFone}[1]{\ensuremath{\mu_{\ensuremath{\mathrm{F}_1}}}}
\newcommand{\meanMacroFone}[1]{\ensuremath{\mu_{\ensuremath{\mathrm{M}\text{-}\mathrm{F}_1}}}}

\newcommand{\ADP}[1]{\( \mathrm{ADP} \)}

\usepackage{pgf}

\NewTableCommand{\heatcell}[1]{%
  \pgfmathparse{int(#1*100)}%
  \SetCell{bg=red!\pgfmathresult}%
}

\definecolor{low}{RGB}{173, 216, 230}   %

\definecolor{high}{RGB}{255, 245, 153} %

\NewTableCommand{\gcreverse}[1]{%
  \pgfmathparse{int(#1*100)}%
  \SetCell{bg=low!\pgfmathresult!high,fg=black,halign=c,valign=m}%
}

\NewTableCommand{\gc}[1]{%
  \pgfmathparse{int((1-#1)*100)}%
  \SetCell{bg=low!\pgfmathresult!high, fg=black, halign=c, valign=m}%
}

\NewTableCommand{\gcrREVERSE}[3]{%
  \pgfmathsetmacro{\norm}{(#1 - #2) / (#3 - #2)}%
  \pgfmathsetmacro{\normclamp}{max(0, min(1, \norm))}%
  \pgfmathparse{int(\normclamp*100)}%
  \SetCell{bg=low!\pgfmathresult!high, fg=black, halign=c, valign=m}%
}

\NewTableCommand{\gcr}[3]{%
  \pgfmathsetmacro{\norm}{(#1 - #2) / (#3 - #2)}%
  \pgfmathsetmacro{\normclamp}{max(0, min(1, \norm))}%
  \pgfmathparse{int(\normclamp*100)}%
  \SetCell{bg=high!\pgfmathresult!low, fg=black, halign=c, valign=m}%
}
  
\usepackage[hidelinks]{hyperref}
\usepackage{orcidlink}

\def\BibTeX{{\rm B\kern-.05em{\sc i\kern-.025em b}\kern-.08em
    T\kern-.1667em\lower.7ex\hbox{E}\kern-.125emX}}
\usepackage{balance}

\begin{document}
\title{\Large \bf \grasp{} - Graph-Based Anomaly Detection Through Self-Supervised Classification}
\author{
      \IEEEauthorblockN{
        Robin Buchta\textsuperscript{\orcidlink{0000-0001-5335-5735}}, 
        Carsten Kleiner\textsuperscript{\orcidlink{0000-0001-9497-0312}},
        Felix Heine\textsuperscript{\orcidlink{0000-0001-7290-6037}},
        Gabi Dreo Rodosek\textsuperscript{\orcidlink{0000-0002-8702-8553}} \\
      }

\thanks{This work has been submitted to the IEEE for possible publication. Copyright may be transferred without notice, after which this version may no longer be accessible.}}

\maketitle

\begin{abstract}
Advanced persistent threat (APT) attacks remain difficult to detect due to their stealth, adaptability, and use of legitimate system components. Provenance-based intrusion detection systems (PIDS) offer a promising defense by capturing detailed relationships between system components and actions. 
However, current PIDS rely on predefined or subset-determined thresholds, which limit detection stability and the ability to detect any anomalous behavior in general. 
Furthermore, related work often neglects the role of process executables, which describe system activity by interacting through a process with files, network components, and other processes.
We introduce \grasp{}, a PIDS based on masked self-supervised classification. 
\grasp{} masks the executable information of processes and learns to infer it from their two-hop provenance graph neighborhood, marking misclassified processes as anomalies. 
It captures behavior patterns for the learned executables without thresholding, making it robust against interference and unknown activities.
Evaluations on the DARPA TC and OpTC datasets demonstrate that \grasp{} consistently detects anomalous behavior, including known attack-related activities, outperforming existing systems. 
Our PIDS identifies all documented attacks on datasets where the behavior of executables is learnable. 
In addition, compared to existing systems, \grasp{} uncovers potentially malicious anomalous behavior not labeled as an attack in the documentation. 
\end{abstract}


\section{Introduction}\label{sec:introduction}
Detecting advanced persistent threat (APT) attacks is a challenging task due to their stealth, persistence, and use of novel techniques~\cite{apt_survey}.  
APT actors use techniques such as ``living-off-the-land'' (LOTL), in which legitimate executables are used to infiltrate the system~\cite{dod_living_off_the_land, apt_survey_buchta}.
Critical systems, including central server infrastructures and industrial environments, represent strategic targets~\cite{dni}. 
To better protect these systems, the focus is not only on defense and external protection, but also on monitoring and detection. 
It is assumed that APT actors are already in the systems or will gain access~\cite{apt_survey_buchta, apt_survey}. 

Therefore, monitoring system behavior and detecting anomalies potentially indicative of attack activity is crucial. 
Detecting these anomalies requires suitable data. 
Provenance data generation has attracted attention for monitoring system behavior, as it provides a fine-grained, comprehensive view of system activities~\cite{sok_collection, pids_survey, graph_detection_survey}.
DARPA has supported the generation of provenance data and the publication of datasets through the Transparent Computing (TC)~\cite{darpa_starc} and Operationally Transparent Cyber (OpTC)~\cite{optc} programs. 
These datasets serve as benchmarks for PIDS~\cite{what_we_talk} and support the evaluation of one-class learning methods, since both normal and malicious behaviors are present.
One-class learning methods are suitable for the use case to detect APT activities, since they do not require knowledge of attack patterns, which are unavailable for APT attacks~\cite{apt_survey_buchta, orthrus, velox}. 
Furthermore, attackers' actions are unpredictable, and even if they are detected, determined attackers will adapt their methods and try again~\cite{apt_survey_buchta}.

However, these datasets also have disadvantages; for example, Liu et al.~\cite{what_we_talk} criticize the lack of detailed attack documentation and the presence of synthetic normal behavior that represents routines. 
This lack of attack documentation complicates evaluation because the ground truth (\gt{}) information is ambiguous. 
As a result, several different \gts{} already exist~\cite{threatrace, orthrus, velox, what_we_talk}. 
The regular, repetitive normal behavior mirrors typical Cyber-Physical System (CPS) operating rhythms~\cite{challenges_buchta, scada_book}, implying that PIDS trained on these datasets generalize better to CPS and server environments than to more diverse end-user systems.

We transform provenance data into graphs that capture dependencies and contextual information among processes, files, and network nodes~\cite{pids_survey, graph_detection_survey}.
Provenance graphs model relationships by encoding individual events along with the context of each node’s neighbors.  
For example, a legitimate system process writes to a file that another process later executes. 
They reveal chained dependencies, enabling the detection of complex attack patterns that conventional detection methods miss, since they do not correlate surrounding information.
Moreover, these relationships are difficult for attackers to mimic, making it hard to hide the actions of an attack~\cite{220315}.

Graph neural networks (GNNs) are advantageous because they leverage graph structure and node features to learn patterns without resource-intensive feature processing~\cite{grahsage, everything_is_connected,graph_detection_survey, fudd}. 
Persistent challenges include high run-to-run instability and difficulty modeling diverse process behaviors. 
Previous methods focus on predicting node~\cite{flash,threatrace,r-caid}, edge types~\cite{threatrace, kairos, orthrus, velox}, or using negative sampling~\cite{shadewatcher, tgb_negsamle, fudd}, which insufficiently capture overall system behavior. 
None of the existing systems can detect attacks and malicious behavior in novel executables with an acceptable number of false positives.
To maintain detection accuracy, these results are post-processed using clustering~\cite{orthrus} or thresholding, either predefined~\cite{flash} or subset-based~\cite{velox}.

Additionally, the role of process executables is either overlooked~\cite{threatrace} or embedded in a sentence embedding for an edge~\cite{flash,orthrus,velox}, even though we consider them primarily responsible for system behavior, since processes interact with files, networks, and other processes. 
We consider processes to be the most important node type and thus focus on classifying their executables to represent the system behavior.

To use the process information for learning system behavior and as a training target for our GNN, we chose masked self-supervised classification.  
With this method, the class, in our case, the executable of a node, is masked, and the model learns to predict it based on its neighborhood. 

Therefore, we guide our research with the following question: 
Is a graph-based anomaly detection system that uses masked self-supervised classification of process executables effective at detecting APT attacks?
Our approach, \grasp{} (GRaph-based Anomaly detection through Self-suPervised classification), uses masked self-supervised classification of process executables to detect anomalies.
Each process is associated with an executable that determines its behavior.
These executables are known during data collection and stored as node attributes within the graph. 
By learning to predict the correct executable based on its neighborhood, deviations indicate anomalous activity, as the node behaves differently from the executable's learned behavior during training. 

In our experiments, depending on the dataset, we detected all reported attacks by identifying unexpected process behavior that we could associate with the attack using \gt{} information.
In addition, we can associate up to 92\,\% of reported false positives with anomalous behavior not documented in the \gt{} information (see Table~\ref{tab:anoamlies}).

Furthermore, we can show that even if we do not detect any of the few attack indicators recorded by \gt{}, we still record significant clusters of anomalies during the attack periods. 
In summary, our contributions are as follows: 

1.) An anomaly detection approach based on classifying process executables for attack detection. 
2.) Demonstration of robust classification that eliminates the need for threshold tuning. 
3.) Time-window-based reporting to support tracking anomalies.  

The \grasp{} source code and supporting materials are available online\footnote{\url{https://github.com/grbaande/grasp}} to support reproducibility and future work. 

The remainder of this paper is organized as follows: 
Section~\ref{sec:threat_model} displays our considered threat model,
Section~\ref{sec:background} explains selected methodical aspects and reviews related work, 
Section~\ref{sec:grasp} details the \grasp{} architecture and methodology, 
Section~\ref{sec:evaluation} presents our evaluation, 
Section~\ref{sec:discussion} discusses findings and implications, 
and Section~\ref{sec:conclusion} concludes with future directions and a summary.

\section{Threat Model}\label{sec:threat_model}
Following prior work~\cite{kairos, unicorn, poirot, holmes, Wang2020, orthrus}, we assume adversaries seek to gain control of and maintain persistence within the target system. 
Hardware-based side channels and attacks that do not use system calls and therefore cannot appear in the logs under consideration are out of scope. 
The system to be monitored is trusted during data collection and model training, excluding data or model poisoning attacks. 
The trusted computing base (operating system, provenance capture, and supporting software) is assumed hardened against manipulation, with provenance logs protected by tamper-evident mechanisms. 

\section{Background}\label{sec:background}
To detect potential APT activities, we rely on anomaly detection on the processes’ execution contexts. 
In the datasets we use, subjects\footnote{Subjects refer to the processes in the provenance graphs used; thus, the terms are interchangeable in this work.} which represent execution contexts through processes~\cite{255010}. 
The execution contexts are determined by the file executed via the execute system call.
The datasets represent provenance graph information generated from audit logs, which provide a detailed record of system behavior~\cite{what_we_talk}, including the processes with their execution contexts, which holds characteristics for communicating with other nodes, such as netflows, files, or other processes. 

We learn the process behavior using the two-hop neighborhood.  
The two-hop neighborhood includes both direct process information and direct and indirect impacts of the target process actions. 
A higher neighborhood level would excessively increase the amount of information to consider and introduce behaviors unrelated to the target node.  
In addition, the available information increases the required storage space. 
Also, the computational resources required to calculate the neighborhoods increase. 
For example, we show how the one- to four-hop neighborhoods increase in Appendix~\ref{app:neighborhoods}. 
The results show that the 1-hop neighborhood provides little contextual information in the chosen setting.
The two-hop neighborhood contains more information than the one-hop neighborhood. 
The 3-hop and 4-hop neighborhoods required significantly more computational resources to compute and use, yet they did not improve detection performance.

However, even in the 2-hop neighborhood, there are nodes with particularly high neighborhood counts despite relatively low mean values, such as \theiaEthree{} with 202,755, which cannot be calculated efficiently. 

To overcome this limitation, we used a sampling strategy based on~\cite{graphsage} to compute even large neighborhoods.
An alternative approach is to reduce the context size for the graph construction. 
We explain the respective concepts in Section~\ref{sec:grasp} and their effects in Subsection~\ref{subsec:hyperparamter}.

\begin{figure}[h]
  \centering
  \includegraphics[width=0.99\columnwidth]{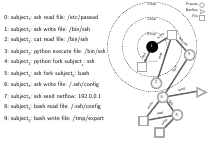}
  \caption{Abstract example of system behavior: From raw provenance logs to a graph with a target node and its 2-hop neighborhood.}
  \label{fig:graph_motivation}
\end{figure}

Figure~\ref{fig:graph_motivation} shows an abstract example based on the data used. 
The Figure shows a graph with ten edges, where the node of the first subject holds as executable \texttt{ssh}. 
In the one-hop neighborhood, it is known that the target process reads the \texttt{/etc/passwd} file and writes the \texttt{/bin/python} file. 
In the two-hop neighborhood, the indirect effects become apparent: the second subject executes the written file and acts as a \texttt{python} executable.
Additionally, the file was read by a process using the \texttt{cat} executable. 
The other edges displayed are outside the two-hop neighborhood. 

As mentioned in Section~\ref{sec:introduction}, there are no learnable examples of APT attacks, and we must account for LOTL techniques.
Therefore, we must detect such attacks without assumptions about the attacks or training examples~\cite{apt_survey_buchta}. 
To achieve this, we use a self-supervised learning approach that learns system behavior without attack information~\cite{ssl}.
We hide one attribute from the input features of our GNN and use it as the target for prediction. 

\begin{figure}[h]
  \centering
  \includegraphics[width=0.99\columnwidth]{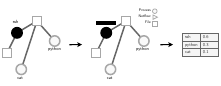}
  \caption{Example of masked self-supervised learning.}
  \label{fig:self-supervised_example}
\end{figure}

Figure~\ref{fig:self-supervised_example} shows an example based on Figure~\ref{fig:graph_motivation}.
In Figure~\ref{fig:self-supervised_example}, the edge types are hidden, and the executables of the process nodes are displayed to focus on self-supervised learning.

On the left, we show the target node (filled black) and its two-hop neighborhood. 
The graph is then transformed by hiding the executable information of the target node, displayed in the middle. 
In our case, this graph is used by a GNN encoder and a Multi-Layer Perceptron (MLP) decoder to determine the true executable of the target node.
The abstract example shows that the target node has a 60\,\% probability of being \texttt{ssh}, a 30\,\% probability of being \texttt{python}, and a 10\,\% probability of being \texttt{cat}.

\subsection{Related Work}
The concept of graph-based anomaly detection via self-supervised classification is already used in various PIDS. 
For example, \orthrus{}~\cite{orthrus} is the most closely related work and uses a self-supervised graph learning method for edge type prediction. 
The authors of \orthrus{} chose a lightweight version of the Temporal Graph Network (TGN) and therefore do not consider the two-hop neighborhood.
Instead, they focus on a node's recent neighbors, e.g., the last 10 actions. 
 The TGN originally maintained a state for each node and updated it with new actions; this memory component is omitted from their architecture to reduce resource consumption. 
Another difference is that we have executable prediction, while they have edge-type prediction. 
Additionally, they rely on validation set-based thresholds and a clustering mechanism for reporting, 
which does not enable real-time application and limits practicality, as the thresholds would need to be determined using test and validation data. 

\orthrus{} and earlier related work, which is less closely related to our work, 
were compared in a recent study by Bilot et al.~\cite{velox}.
Although the latest PIDS deliver excellent results, 
they have limited practical relevance~\cite{velox, are_we_there_yet}. 
The authors identify nine shortcomings, which we adopt and enrich. We list the shortcomings used in Subsection~\ref{subsec:shortcomings}. 
The shortcomings guide our architecture and evaluation. 
Another closely related system is \velox{}~\cite{velox}, which addresses the previously identified shortcomings.
\velox{} performs better than \orthrus{} and addresses all nine shortcomings, but it also fails to address our four additional identified shortcomings. 
\velox{} uses the same self-supervised learning method as \orthrus{}, but on a simpler architecture. 
The architecture consists of three MLPs: one encodes the source node, another encodes the destination, and the third serves as the decoder.  
\velox{} does not use any additional information for neighboring actions. 
This setup enables real-time operation. 
At \velox{}, we see shortcomings, such as high instability, overlooking obviously anomalous behavior, omitting analysis of false-positive reports, using unrealistic training and validation splits, and the results are not reliable reproducible.
Recent systems, especially \velox{} and \orthrus{}, are not robust. Hence, the best evaluation is significantly better than the average performance across several runs; see Appendix~\ref{app:velox}, where we show the results of the comparison from~\cite{velox}, including those of \grasp{}. 
For example, the reported precision values for \cadetsEthree{} range from 0.44 to 1.00.

Furthermore, the systems are sensitive to the chosen threshold. 
An experiment (Appendix~\ref{app:reeval}) shows that replacing the maximum threshold with the 99.9th or 99th percentile significantly reduces reproducibility, because filtering depends on single-outlier event thresholds. 
\orthrus{} yields more stable report counts due to its clustering layer, yet still fails to detect the attacks. 
We also attempted to reproduce the results, both using default values and the hyperparameters tuned for each dataset as published by the authors; however, we were unable to do so. 
In our runs, the detection performance of related works were not reproducible (see Appendix~\ref{app:reeval}). 

\velox{} and \orthrus{} use edge classification as a self-supervised learning method to distinguish between 10 event types. 
Both attack and normal behaviors are more complex than can be captured by edge types alone, as multiple edges can be equally valid at a given process point.
Additionally, the method’s success depends heavily on thresholding, whether via clustering~\cite{orthrus} or thresholds determined from a predefined validation dataset~\cite{velox}.
The reports are at the node level and can be processes, files, or network connections. 
We focus solely on processes, as attacks are typically carried out through them. While files and network connections are less indicative on their own, they provide valuable context for classifying process behavior.

\subsection{Shortcomings}\label{subsec:shortcomings}
Our development of \grasp{} is guided by the shortcomings identified by Bilot et al.~\cite{velox} and evaluated on a measurable subset of these, along with our four new shortcomings (underlined in the following enumeration).
We grouped them into two categories\footnote{Due to our grouping, the numbering of the SCs differs from that of~\cite{velox}.}.

\begin{table}[!h]
    \centering
    \caption{Shortcomings addressed by systems. \cmark = addressed, \cstar = improves rel. work.\label{tab:shortcomings}}
    \begin{tblr}{
        width=\columnwidth,
        colspec={l*{8}{X[c]}},
        cells={font=\scriptsize},
        row{1}={valign=m,c,font=\bfseries\footnotesize},
        hline{2-Y} = {1-Z}{0.2pt, dashed},
        vline{2-Y} = {2-Z}{0.2pt, dashed},
        hline{2} = {0.5pt, solid},
        vline{1,2,Z} = {0.5pt, solid},
        hline{1,Z} = {0.5pt, solid},
        hline{1} = {1}{0pt},   
        vline{1} = {1}{0pt},   
        vline{6} = {0.5pt, solid},   
        colsep=2pt,
        rowsep=1pt,
        }
        {} &
        \SetCell[c=4]{c}{Methodology\\Issues} & & & & 
        \SetCell[c=4]{c}Evaluation Gaps               \\
        \textbf{System}                 & {1}    & {2}    & {3}    & {4}    & {5}    & {6}    & {7}     & {8}    \\
        \textcolor{orthrus}{\orthrus{}} & \cmark &        &        &        &        &        &        & \cmark  \\
        \textcolor{velox}{\velox{}}     & \cmark & \cmark & \cmark &        & \cmark &        &        & \cmark  \\
        \textcolor{grasp}{\grasp{}}     & \cmark & \cstar & \cmark & \cmark & \cmark & \cmark & \cmark & \cstar  \\
    \end{tblr}
    
\end{table}

Our first category, methodology issues, lists shortcomings in the design of the detection system:

\noindent\textbf{SC\,1 Insufficient Detection Granularity:} 
Most systems detect attacks at coarse levels, overwhelming analysts and limiting practicality~\cite{velox}.

\noindent\textbf{SC\,2 Impractical Thresholding Methods:} 
Systems use impractical thresholding, relying on fixed or data-dependent methods that are hard to adapt and risk data snooping~\cite{velox}.

\noindent\textbf{SC\,3 Featurization Methods Trained on Test Data:} 
Text embeddings often struggle with unseen words, requiring either data snooping or unclear zero-vector substitutions~\cite{velox}.

\noindent{\underline{\textbf{SC\,4 Missing Anomaly Detection Capacity:}}} 
Recent related work (e.g.,~\cite{orthrus, velox}) reports nearly perfect attack detection, leaving no anomalies flagged, even though datasets like Cadets E3 include changes in normal behavior~\cite{what_we_talk}, that are not part of the attacks.

The second category, evaluation gaps, highlights shortcomings in the methods used for assessment:

\noindent{\textbf{SC\,5 Not Measuring Instability:}}
Many studies overlook predictive instability in PIDS's, where random factors such as weight initialization can yield inconsistent results even under identical settings~\cite{velox}.

\noindent{\underline{\textbf{SC\,6 Missing Analysis of False Positives:}}}
False positives in anomaly-detection methods are not analyzed; they may indicate undocumented attacks or changes in normal behavior.

\noindent{\underline{\textbf{SC\,7 Selected Training and Validation Splits:}}}
A shortcoming of related work is a dataset-customized selection of training and validation periods. Other validation periods can significantly influence the results, especially when using a non-robust threshold method, such as the maximum loss from a period. Experiments on this are included in Appendix~\ref{app:reeval}. 

\noindent{\underline{\textbf{SC\,8 Reproducibility:}}}
The excellent results reported in the related work cannot be reproduced using robust metrics averaged across multiple runs. 
In their repository, the authors explain that dataset-specific hyperparameters and lucky runs are necessary to achieve these results\footnote{\url{https://github.com/ubc-provenance/PIDSMaker/issues/27}}. 
We have included our reevaluation of the related work in the Appendix~\ref{app:reeval}.

Since \velox{} has already compared itself with~\cite{sigl,threatrace,NODLINK,magic,kairos,flash,r-caid,orthrus}, 
and \velox{} performs better across all areas of their shortcomings, we limit our comparison to the two most recent systems, \orthrus{} and \velox{}.
Table~\ref{tab:shortcomings} compares the shortcomings of \orthrus{}, \velox{}, and \grasp{}.
With \grasp{}, we address all eight shortcomings.

We also improved on SC\,2, which is marked with a star compared to the related work  \velox{}, as  \velox{} uses thresholding based on a validation dataset, whereas \grasp{} does not require a threshold.
In addition, we will examine the system's ability to detect anomalies. 
\orthrus{} addresses just SC\,1 as also mentioned in~\cite{velox} within the first category. 
\orthrus{} operates at node-level granularity, which is practical. 
Bilot et al.~\cite{velox} also introduced a new metric, Attack Detection Performance (ADP), which we adopt and detail in Section~\ref{sec:evaluation}.
\velox{} addresses lacks in SC\,4, SC\,6, and SC\,7. 
Because, e.g., for SC\,4, they report a precision of 1.00 on the \cadetsEthree{} dataset, despite changing normal behavior~\cite{what_we_talk} after the first attack. 
We list the full comparison in Appendix~\ref{app:velox}. 
Furthermore, as mentioned before, they used custom training and validation splits, which is critical, because the threshold is based on the chosen validation data. 
In the validation data, only the highest loss value is considered, making the methodology dependent on individual outliers. 
All systems comply with SC\,8; to be fair, we must note that we were able to produce similar results in some cases, though not consistently across multiple runs, as the authors point out. 
We surpass this, however, as we deliver reproducible results across multiple runs. 
We have verified through experiments in which students were able to reproduce our results.

\section{Method: \grasp{} Approach}\label{sec:grasp} 
\begin{figure*}
  \centering
  \includegraphics[width=0.8\textwidth]{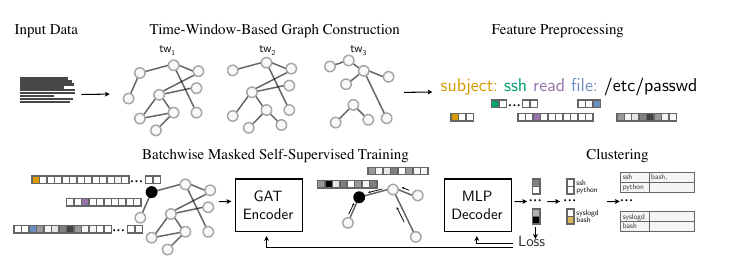}
  \caption{Training pipeline of \grasp{}.}
  \label{fig:training}
\end{figure*}

\begin{figure}[h!]
    \centering
    \includegraphics[width=1.0\columnwidth]{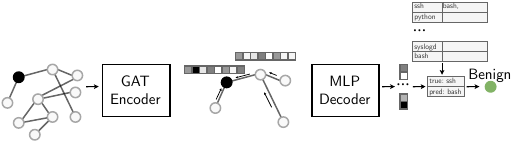}
    \caption{Inference pipeline of \grasp{}.}
    \label{fig:inference}
\end{figure}

Figures~\ref{fig:training} and~\ref{fig:inference} provide an overview of \grasp{}’s training and inference pipelines. 
Training begins with provenance data collected from various sources.  
In the examples in this paper, the data collection includes \cadets{}, \theia{}, \Clearscope{}, and \optc{}, covering Linux, Android, and Windows systems. 
For each source separately, we segment the data into sliding windows, and each window is converted into a graph.

We preprocess these graphs into GNN-compatible representations through feature extraction and encoding. 
Node and edge attributes are one-hot encoded, while paths and IP addresses are represented using transformer-based autoencoder embeddings. 
The resulting graphs are fed into a GNN-based anomaly detection model composed of a Graph Attention Network~\cite{gat} (GAT) 
encoder and an MLP decoder.
Each target node and its neighborhood are processed to generate embeddings, which the decoder classifies to estimate the likelihood of known executables.
Training optimizes the classification loss iteratively, and misclassified nodes after training are clustered to account for indistinguishable behaviors. 

Our clustering approach uses misclassifications after the model's last epoch on the training data to account for indistinguishable entries that behave identically to the model. 
The clusters list these benign permutations for inference.
During inference, provenance data from the monitored system is processed using the same preprocessing 
and feature extraction steps as in training, but in inference mode. 
The training provides the encoding and the trained autoencoder for the paths and IP addresses. 
We generate sliding-window graphs and pass them into the trained GNN model. 

Each target node is classified by the decoder, which outputs probabilities for known executables. 
Nodes (process behaviors) that are misclassified and do not fit into any stored cluster are flagged as anomalies.
Further technical details are described in the Appendix~\ref{app:dl_methodology} and the implementation is available within our GitHub repository.

\subsection{Input Data}\label{subsec:input_data}
Our system can use provenance data from different data collectors, as demonstrated by the DARPA data~\cite{darpa_starc,optc}.
Since \grasp{} is based on process-executable classification, the data must contain distinguishable executables and corresponding training data for each. 
We determine whether the executables are distinguishable and learnable by collecting macro- (\MacroFone{}) and weighted-F1 (\Fone{}) scores during training; see Section~\ref{sec:evaluation}. 

In addition, training data must reflect normal behavior to be used for anomaly detection. 
Attack patterns present in the training dataset cannot be identified later as attacks, because we try to adapt our model closely to the training data to respond sensitively to changes.

\subsection{Time-Window-Based Graph Construction}
Our system uses the same data and the same filtering procedures as \orthrus{}~\cite{orthrus} and \velox{}~\cite{velox}. We list the used components of the provenance data in Appendix~\ref{app:graph_info}.
This is essential to ensure comparability with related work. We agree that the filtered nodetypes, edgetypes, and attributes are the most important for representing the system's behavior\footnote{All further links to the data are available in the documentation \url{https://ubc-provenance.github.io/PIDSMaker/}}.  
The raw logs of the datasets used are also publicly available\footnote{The \darpa{} TC Datasets, \cadets{}, \theia{}, and \Clearscope{} \url{https://github.com/darpa-i2o/Transparent-Computing}. 
The \optc{} datasets are under \url{https://github.com/FiveDirections/OpTC-data} available}.
We differ from related work by using an time‑window‑based graph construction controlled by two hyperparameters, context and step size, and by using undirected graphs to better capture process interactions. 
For example, it is equally relevant to a process whether something is read or written in a neighborhood, as one action, e.g, could load harmful information, while the other could cause sensitive information to leak.
The main memory limits the context size, and the step size affects the computation time but yields fine-grained alarms. 
We show the trade-offs in Section~\ref{sec:evaluation}.

\subsection{Feature Preprocessing}
In our case, most attributes can be one-hot encoded, which requires knowledge of the number of possible categories for each feature. 
The node type, which can take the values Subject, File, or Netflow. 
The edge type, which has 10 values, is shown in Appendix~\ref{app:graph_info}. 
The executable information provided with the Subject node has a variable length, determined by the number of distinct executables in the training data. 
During inference, an unknown executable could appear due to new normal behavior or malicious actions.
In this case, we model the corresponding feature vector as a vector of zeros. 

The location information, including path and IP address details, is encoded using a transformer-based autoencoder~\cite{attention_is_all_you_need}, implemented in PyTorch. 
To do this, a location is split into character-based segments, and positional encoding is applied. 
We have limited the characters to numbers, ASCII letters, punctuation, and white space, totaling 100 characters. 
The encoder's output dimension determines the subsequent machine-learning-capable representation. 
This approach prevents data snooping and requires no out-of-vocabulary defaults.

\subsection{Masked Self-Supervised Training}
Our Training uses sliding-window graphs with batches.
The batch size corresponds to the number of processes in the sliding window. 
Our GAT encoder receives the masked target process node and its two-hop neighborhood, along with the bidirectional edges connecting them. 
Using this information, a representation of the process is created via message passing~\cite{message_passing}, incorporating information from the two-hop neighborhood and from itself.  

The target node's embedding size is a hyperparameter and serves as input to the MLP decoder, which predicts the target process's executable. 
The deviation between the predicted and true values is measured using cross-entropy loss, and backpropagation through the decoder and encoder is initiated.
The training runs across batches within each sliding window and across all sliding windows over multiple epochs. 
The particular specifications for batch size and number of epochs are hyperparameters. 

\subsection{Clustering}
We use misclassifications after the model’s final training epoch to identify entries whose behavior is indistinguishable.
If the prediction differs from the monitored value, we cluster them by creating an entry for the actual executable with the prediction, as illustrated in Figure~\ref{fig:training}. 
Each executable in the training data has an entry, and each entry contains a list of these mixups.
This applies to executables that are very similar in behavior and to those that are significantly underrepresented. 
The behavior and the distinguishability of the executables are monitored, as mentioned in Subsection~\ref{subsec:input_data}.

Through self-supervised learning and clustering, the model captures the normal behavior observed in the data. 
Inference on the training data then shows that all processes are classified as benign, allowing any deviating behavior in the inference on the test data to be flagged as malignant.
Retraining is possible by extending feature preprocessing. 
Continuous training is also possible, unless new executables are to be added to the model, in which case retraining would be required, since the multi-class classification method used requires the number of possible outcomes to be determined in advance. 
We determine this based on the training data, which makes it feasible in practice; however, it requires the training data to be available in advance.

\subsection{Inference}
The inference process in \grasp{} is identical to the training process, except that no loss is calculated and the existing preprocessing is used. 
Therefore, the encoder and decoder weights are not further modified, and the clustering is not changed. 
The determined values are only compared with the clustering. 
Each subject node is processed using the MLP decoder's softmax output to produce a probability distribution. 
The highest value determines the label of the target node. 

The prediction is compared with the monitored value, yielding three possible outcomes: (1) a matching prediction, indicating benign behavior; (2) a misclassification covered by clustering, also benign; and (3) a case where neither condition holds, indicating malicious or abnormal activity.

\subsection{Shortcomings Addressed} \label{subsec:sc_addressed}
Our architecture addresses all the methodological shortcomings listed in Table~\ref{tab:shortcomings}.
Our reports are generated as time-windowed node-level graphs, where the alarm refers to the target node and the neighborhood is included solely for contextualization. 
\orthrus{} is based on a clustering algorithm that clusters nodes without any time-related information. 
\velox{} could report in much greater detail since it considers each log line individually, but reports at the node level despite making edge predictions. 
By not using a threshold method, we have addressed SC\,2 more effectively than the related work, which still relies on a threshold. 
In addition, a thresholding method or result clustering can be applied to our alarms.

\orthrus{} fails to satisfy the requirement due to its thresholding mechanism, which depends on validation and test sets. 
Similarly, \velox{} relies on a validation set, which substantially contributes to its performance. An anomalous event in the validation data could elevate the threshold excessively, thereby compromising detection entirely.

By implementing a transformer-based autoencoder and one-hot encoding, we avoid data snooping and do not need default values for out-of-vocabulary locations, and therefore, we treat SC\,3.
The only out-of-vocabulary defaults occur when an unknown executable appears in the two-hop neighborhood. 
In this case, the feature vector for the executable is set to zero. 

\grasp{} flags any unusual behavior for further analysis, which enables us to target SC\,4.
Whereas \velox{}, in particular, rarely produces false positives in terms of attack detection and therefore does not recognize further behavioral changes that take place in the monitored systems.  

The sliding window-based implementation, with its context, step, and batch size, provides us with control elements that allow us to trade off between main memory, processing speed, and more detailed reporting.

\section{Evaluation}\label{sec:evaluation}
\grasp{} is primarily an anomaly detection system that we use to detect attacks.
We have therefore divided our evaluation into different parts. 
First, we explain the evaluation metrics, the dataset, the baseline systems, and the hardware configuration. 
This is followed by a detection performance evaluation and an example in-depth analysis.
Finally, we examine the impact of hyperparameters and conduct a LOTL attack study.

\subsection{Metrics}
We agree with previous work~\cite {orthrus,threatrace,flash} that precision and recall are both important. 
Still, we also agree with Dong et al.~\cite{are_we_there_yet} that it is more important to detect all attacks (high attack recall) than to detect every single attack step (high attack step recall). 
Furthermore, it is difficult to identify every step of an attack from the data to create an accurate \gt{}~\cite{orthrus}. 

In addition, the datasets also contain undocumented behavior, such as changes in the normal behavior, during the test periods. 
This is demonstrated by the dataset analysis conducted by Liu et al.~\cite{what_we_talk}, which also explains our findings, as the amount of unseen behavior added during test periods varies across datasets. 
The unseen behavior can be interpreted as either an undocumented attack or a changed normal behavior. 
However, we consider these reports anomalous and analyze entries that clearly indicate unknown behavior separately. 

Related work~\cite{velox} builds on the insight that attacks can be reconstructed from only a few correctly identified nodes~\cite{277080, Liu2018TowardsAT, 9833632}.
In the \velox{} evaluation, a single node, regardless of type, is sufficient to mark an attack as detected.

With \( D(p) \) calculating the percentage of attacks detected with a precision of \( p \), defined by: 
\begin{equation*}\label{eq:Dp}
D(p) = \frac{\lvert \{\, A_i \mid A_i \cap R(p) \neq \emptyset \,\} \rvert}{k}
\end{equation*}
where \( A_i \) is the node set of attack \( i \)  among the total of \( k \)  attacks, and \( R(p) \) are the reported nodes with precision \( p \). 

Our approach does not use a threshold; therefore, we report just the \( D(p) \). 
Since \( D(p) \) indicates recall over the detected attacks, we refer to this metric as \attackrecall{} in this paper. 

We contend that achieving perfect precision in attack detection is unrealistic for anomaly detection, as it would dismiss all other deviations thereby limiting its practical applicability in operational settings. 
Furthermore, perfect precision would mean recognizing all novel normal behavior patterns as benign, which is impossible for methods based on anomaly detection without fine-tuned result-filtering techniques, such as threshold optimization or clustering, because no training samples reflect this as normal. 

We treat threshold determination as part of the system~\cite{orthrus, velox}, and to evaluate its practical relevance, we do not integrate values across different thresholds\footnote{Except in our reevaluation study, in which we use all of the authors' metrics (Appendic~\ref{app:reeval}).}. 
In addition, we measure the number of anomalies caused by undocumented anomalous behavior.
We refer to the findings of Liu et al.~\cite{what_we_talk}, who state that unknown executables are anomalous by definition. 
We measure this by reporting the percentage of unknown executables reported. 
\grasp{} includes the detection of unknown executables in its system design, whereas the related work almost entirely fails to report these activities. 
We determine the value to compare against existing systems and measure the extent of our deviation in the LOTL attack study. 

We evaluate \grasp{}’s training quality using macro-F1 (\MacroFone{}) and weighted-F1 (\Fone{}), which indicate its ability to learn executables. 
We collect the values from the last epoch on the training data before clustering.

\subsection{Datasets}
We use the provenance data from DARPA's TC and OpTC programs as datasets. 
We use a selection of data from \cadets{}, \theia{}, and \Clearscope{} recorders, as well as OpTC-Hosts 51, 201, and 501. 
Liu et al. \cite{what_we_talk} examined the data in more detail and found them suitable for evaluating PIDS.
Although the datasets have their weaknesses, they remain the standard given the availability of suitable datasets~\cite{what_we_talk,velox,orthrus}.  
We do not consider the sometimes criticized recurring normal behavior a weakness, since it reflects a representative operational environment, from centralized infrastructure to CPS, enabling the learning of normal patterns that would otherwise be unattainable given the impossibility of anticipating all novel behaviors.
The datasets include normal behavior, which allows anomaly detection methods to be developed and evaluated through the attacks carried out. 
We attached additional information about the datasets in Appendix~\ref{app:dataset_stats}.

In related work, different training/validation and test splits of the dataset were used. 
To ensure comparability with values from related work, particularly the overview work by Bilot et al.~\cite{velox}, and to meet the requirements of~\cite{are_we_there_yet}, we evaluate our system across different splits (see Appendix~\ref{app:splits}) and compare it with the latest related work. 
Our evaluation scenario uses historical data up to a given date as training data, and the remaining data for inference. 

\subsection{Systems for Comparison} 
The systems we compare ourselves to are \orthrus{}~\cite{orthrus} and \velox{}~\cite{velox}, as they are the latest and best-performing PIDS. 
The systems themselves have already been extensively compared with previous related work and perform significantly better. 
In addition, the systems are included in a framework\footnote{\url{https://github.com/ubc-provenance/PIDSMaker}}, and we can use their code. 
To compare ourselves with the previous systems, we run those with the default configuration on our development server. 
We compare ourselves with \orthrus{} and \velox{} on our splits.
This comparison is limited to the DARPA TC datasets from the third engagement and the OpTC datasets\footnote{For the OpTC datasets, we had to use shortened training periods (applied to all systems) because the data volumes exceeded our main memory capacity. 
For engagement five, we were unable to identify any feasible splits that would yield insights, as the datasets differ significantly in size.}. 
To compare ourselves with the systems~\cite{sigl,threatrace,NODLINK,magic,kairos,flash,r-caid} and on the datasets from the fifth engagement, we adopted the original evaluation by Bilot et al.~\cite{velox} and supplemented our results. 
We show the results in Appendix~\ref{app:velox}.

\subsubsection*{Hardware Setup}
We performed our evaluation on a server running Debian 13, equipped with an AMD Ryzen 9 7950X 16-Core Processor, 128\,GB of RAM, and an NVIDIA RTX 6000 Ada with 48\,GB of memory. We also use zram to compress the main memory and enable the use of larger datasets.

\subsection{Detection Performance}
Tables~\ref{tab:e3} and~\ref {tab:optc} present the mean results over five runs for the \attackrecall{} (\meanattackrecall{}), the coefficient of variation of \attackrecall{} (\cvattackrecall{}), the average number of alarms (\meana{}), the average detection rate of anomalies based on unseen executables(\meane{}), and the average number of true positives (\meanh{}).

We have listed important hyperparameters used in Appendix~\ref{app:splits} and the splits used in Appendix~\ref{app:parameters}.

As shown in Tables~\ref{tab:e3} and~\ref {tab:optc}, \grasp{} demonstrates superior and more consistent attack detection compared to the existing systems. 
However, this improvement comes at the cost of more alarms. 
Even though \orthrus{} and \velox{} have a high attack recall within the \optcsecond{} dataset, this is based on only one true positive, which in this case points to a file that does not indicate any behavior. 

The comparison systems apply a thresholding mechanism to reduce the alarms, but this suppression is excessive, resulting in the omission of nearly all other anomalous behavior. 
Consequently, events involving unseen executables are largely undetected.
Only in the Theia E3 dataset does \velox{} deliver a \attackrecall{} of 1.0 and more than one or two \hits{} with fewer alarms. 
However, it overlooks 60\,\% of the obvious anomalies.

We attribute this to the observation that \velox{} and \orthrus{} exceed their result filtering, with threshold optimizing and clustering methods, for only a limited number of event types, primarily open, create, clone, execute, and terminate events, depending on the dataset.
This threshold setting aims to minimize false alarms, but excludes events that should be considered anomalous.
In contrast, \grasp{} achieves higher \hits{}, indicating a broader detection coverage. 

We also observe that the thresholding method is bound to single outliers in the validation data. 
Thus, we performed an additional experiment in which we varied the validation data by excluding the top 0.1\,\% and 1\,\% of abnormal events. 
The results confirm our assumption (see Appendix~\ref{app:reeval}).

None of the systems detected Clearscope E3 attacks. 
The dataset also stands out in the evaluation by Bilot et al.~\cite{velox}, which we listed in Table~\ref{tab:full_tblr} in Appendix~\ref{app:velox}. 
For \grasp{}, we can explain the behavior because it did not learn the executable behavior during training, and the \MacroFone{} score for the training data averages 0.02. 
In contrast, for Theia and Cadets, the \MacroFone{} scores are 0.22 and 0.43.

\begin{table}[t]
  \centering  
  \caption{Detection performance comparison: \orthrus{}, \velox{}, and \grasp{} on third-engagement datasets.\label{tab:e3}}

  \begin{tblr}{
    width=\linewidth,
    colspec={|X[1.3,l]|*{5}{X[c]}|},
    row{1}={font=\footnotesize\bfseries},
    cells={font=\footnotesize},
    hlines = {0.2pt, dashed},
    vlines = {0.2pt, dashed},
    hline{1,Z} = {0.5pt, solid},
    hline{2}   = {0.5pt, solid},
    hline{3,6,7,10,11}   = {0.5pt, solid},
    vline{1,2,Z} = {0.5pt, solid},
    vline{8}   = {0.5pt, solid},
    rowsep=1pt,
  }
  \textbf{System} & \meanattackrecall{} & \cvattackrecall{}  & \meanalarms{} & \meane{} & \meanh{}\\
  \SetCell[c=6]{c}\cadetsfirstbox{}\,\cadetsEthree{} \\
\textcolor{orthrus}{\orthrus{}} & \gc{1.0}1.00 & 0.00             & \gcrREVERSE{9.6}{9}{446.8}9.6      & 0.00 &   \gcr{1.4}{0.}{21}1.4\\
  \textcolor{velox}{\velox{}}     & \gc{0.45}0.45 &  0.22       &   \gcrREVERSE{9.4}{9}{446.8}9.4    & 0.05 &  \gcr{1.4}{0}{21}1.4\\
  \textcolor{grasp}{\grasp{}}     &    \gc{1.0}1.00   & 0.00     & \gcrREVERSE{446.8}{9}{446.8}446.8    & 1.00 & \gcr{21}{0}{21}21.0\\
  \SetCell[c=6]{c}\theiafirstbox{}\,\theiaEthree{} \\
  \textcolor{orthrus}{\orthrus{}} & \gc{0.0}0.00 & 0.00  & \gcrREVERSE{25}{0}{372.2}25.0 & 0.35 & \gcr{0.0}{0}{62.2}0.0\\
  \textcolor{velox}{\velox{}}     & \gc{1.0}1.00 & 0.00 & \gcrREVERSE{131}{0}{372.2}131.0 & 0.40 & \gcr{35.6}{0}{62.2}35.6\\
  \textcolor{grasp}{\grasp{}}     & \gc{1.0}1.00 & 0.00 & \gcrREVERSE{372.2}{0}{372.2}372.2 & 1.00 & \gcr{62.2}{0}{62.2}62.2\\
  \SetCell[c=6]{c}\clearscopefirstbox{}\,\ClearscopeEthree{} \\
  \textcolor{orthrus}{\orthrus{}} & \gc{0.0}0.00 & 0.00 & \gcrREVERSE{10}{0}{1577.2}10.0 & 0.00 & \gc{0.0}0.0\\
  \textcolor{velox}{\velox{}}     & \gc{0.0}0.00 & 0.00 & \gcrREVERSE{2}{0}{1577.2}2.0 & 0.00 & \gc{0.0}0.0\\
  \textcolor{grasp}{\grasp{}}     & \gc{0.0}0.00 & 0.00 & \gcrREVERSE{1577.2}{0}{1577.2}1577.2 & 1.00& \gc{0.0}0.0\\
  \end{tblr}
  
\end{table}

\begin{table}[t]
  \centering
    \caption{Detection performance comparison: \orthrus{}, \velox{}, and \grasp{} on OpTC datasets.\label{tab:optc}}

  \begin{tblr}{
    width=\linewidth,
    colspec={|X[1.3,l]|*{5}{X[c]}|},
    row{1}={font=\footnotesize\bfseries},
    cells={font=\footnotesize},
    hlines = {0.2pt, dashed},
    vlines = {0.2pt, dashed},
    hline{1,Z} = {0.5pt, solid},
    hline{2}   = {0.5pt, solid},
    hline{3,6,7,10,11}   = {0.5pt, solid},
    vline{1,2,Z} = {0.5pt, solid},
    vline{8}   = {0.5pt, solid},
    rowsep=1pt,
  }
  \textbf{System} & \meanattackrecall{} & \cvattackrecall{}  & \meanalarms{} & \meane{}  & \meanh{}\\
  \SetCell[c=6]{c}\optcfirstbox{}{}\,\optcfirst{} \\
\textcolor{orthrus}{\orthrus{}} & \gc{1.0}1.00 & 0.00          & \gcrREVERSE{5}{3}{115.0}5.2      & 0.00 & \gcr{1.0}{0}{319.2}1.0\\
  \textcolor{velox}{\velox{}}     & \gc{0.0}0.00 & 0.00       &   \gcrREVERSE{3.7}{3}{115.0}3.6   & 0.00 & \gcr{0.0}{0}{39.2}0.0 \\
    \textcolor{grasp}{\grasp{}}     &    \gc{1.0}1.00   & 0.00      & \gcrREVERSE{115.0}{3}{115.0}115.0    & 1.00 & \gcr{39.2}{0}{39.2}39.2\\
  
  \SetCell[c=6]{c}\optcsecondbox{}\,\optcsecond{} \\
  \textcolor{orthrus}{\orthrus{}} & \gc{1.0}1.00 & 0.00  & \gcrREVERSE{4}{4}{427.6}4.0 & 0.00 & \gcr{1.0}{0}{9.8}1.0\\
  \textcolor{velox}{\velox{}}     & \gc{0.8}0.80 & 0.50  & \gcrREVERSE{6.2}{4}{427.6}7.0 & 0.00 & \gcr{1.0}{0}{9.8}1.0\\
  \textcolor{grasp}{\grasp{}}     & \gc{1.0}1.00 & 0.00  & \gcrREVERSE{427.6}{4}{427.6}427.6 & 1.00 & \gcr{9.8}{0}{9.8}9.8\\
  
  \SetCell[c=6]{c}\optcthirdbox{}\,\optcthrid{} \\
  \textcolor{orthrus}{\orthrus{}} & \gc{1.0}1.00 & 0.00  & \gcrREVERSE{4}{1}{204.0}4.0 & 0.00 & \gcr{2.0}{0}{23.0}2.0\\
  \textcolor{velox}{\velox{}}     & \gc{0.0}0.00 & 0.00  & \gcrREVERSE{1}{1}{204.0}0.6 & 0.00 & \gcr{1.5}{0}{23.0}1.5\\
  \textcolor{grasp}{\grasp{}}     & \gc{1.0}1.00 & 0.00  & \gcrREVERSE{204.0}{1}{204.0}204.0 & 1.00 & \gcr{23.0}{0}{23.0}23.0\\
  
  \end{tblr}
\end{table}

Table~\ref{tab:e5} shows \grasp{}'s results for the fifth engagement.
We detect all attacks with a low number of false positives, as well as all other anomalous events that we labeled in the Cadets E5 dataset. 
Our multiclass classification also achieves an \MacroFone{} score of 0.43 on the training set, demonstrating that \grasp{} can learn normal behavior patterns. Due to the class imbalance, the \Fone{} score is mostly between 0.97 and 0.99.

With the Theia E5 dataset, our \attackrecall{} metric is 0.00 for the first time with Theia E5. 
We will investigate this in more detail in Subsection~\ref{subsec:detailed_investigation}.

Although the results on Clearscope E3 show that \grasp{} does not work on this kind of data, those on Clearscope~E5 are considered successful, as \grasp{} reports the attacks and anomalous behavior.
As with the Clearscope E3 dataset, we conclude that the normal behavior of Android datasets is difficult to learn and that clustering therefore accounts for the majority of predictions. 
Conspicuous attacks primarily explain the results on Clearscope~E5.

\begin{table}[t]
  \centering
  
  \caption{Detection performance: \grasp{} on fifth engagement datasets.\label{tab:e5}}
  
  \begin{tblr}{
    width=\linewidth,
    colspec={|X[2.0,l]|*{5}{X[c]}|},
    row{1}={valign=m,c,font=\bfseries},
    cells={font=\scriptsize},
    hlines = {0.2pt, dashed},
    vlines = {0.2pt, dashed},
    hline{1,2,Z} = {0.5pt, solid},
    vline{1,2,Z} = {0.5pt, solid},
    rowsep=1pt,
  }
  Dataset & \meanattackrecall{} & \cvattackrecall{} & \meanalarms{} & \meane{} & \meanh{}\\
  \cadetssecondbox{}\,\cadetsEfive{}     &    1.00   & 0.00     & 164.6   & 1.00 & 15.6\\
  \theiasecondbox{}\,\theiaEfive{}     & 0.00 & 0.00 & 160.8  & 1.00 & 0.0\\
  \clearscopesecondbox{}\,\ClearscopeEfive{}     & 1.00 & 0.00 & 108.6  &1.00 & 30.0 \\
  \end{tblr}
  
\end{table}

Table~\ref{tab:anoamlies} presents the analyst's view.
We explain the Table exemplarily using the Theia E3 dataset. 
\grasp{} investigated 162,191 processes and reported 372 nodes. 
This represents a reduction of over 99\,\%. 
In addition, 147 of the reports can be attributed to novel executables that are clearly suspicious. 
This is a further 39\,\% reduction. 
These 147 processes can refer to both attacks and normal behavior that \grasp{} could not learn because they were not present in the training data. 

Our objective is to adhere strictly to the training data and report all deviations. 
Consequently, the analyst must review 147 results and an additional 225 events, instead of over 160\,k alarms without \grasp{}, thereby mitigating alert fatigue. 
Furthermore, our additional alarms do not contribute to alert fatigue, as they report unusual behavior rather than obvious normal behavior.
In our scenario with systems such as \orthrus{} and \velox{}, attacks can only be partially detected, and other unusual behavior is overlooked, which should be considered as malicious~\cite{what_we_talk}.
Furthermore, alarm filtering techniques, as seen in the related work, could be applied.

\begin{table}[t]
\centering
\caption{Evaluation of novel behavior for \grasp{}.\label{tab:anoamlies}}

\begin{tblr}{
  cells={font=\scriptsize},
  width=\linewidth,
  colspec = {lX[r]*{4}{X[r]}},
  row{1} = {valign=m,c,font=\scriptsize\bfseries},
  colsep=2pt,
  rowsep=1pt,
  hlines = {0.2pt, dashed},
  vlines = {0.2pt, dashed},
  hline{1,2,Z} = {0.5pt, solid},
  vline{1,2,Z} = {0.5pt, solid},
  rowsep=1pt,
}
Dataset & {\#~processes} & {\% of alarms}  & \#~alarms & {\# novel processes}  & {\%~of alarms} \\
 \cadetsfirstbox{}\,\cadetsEthree{}     & 149,421 & 0.33 & 446.8& 71 & 15.89      \\
 \theiafirstbox{}\,\theiaEthree{}        & 162,191 & 0.23 &  372.2& 147 & 39.49      \\
 \clearscopefirstbox{}\,\ClearscopeEthree{}  & 12,275 & 12.85 & 1577.2 &  16 & 1.01    \\
 \optcfirstbox{}\,\optcfirst{}{}       & 3,694 & 3.11 & 115.0 & 106 & 92.17     \\
 \optcsecondbox{}\,\optcsecond{}{}        & 19,095 & 2.24 & 427.6 & 273 & 63.84      \\
 \optcthirdbox{}\,\optcthrid{}{}        & 10,931 & 1.87 & 204.0 & 86 & 42.16     \\
 \cadetssecondbox{}\,\cadetsEfive{}      & 2,151,089 & 0.01 & 164.6  & 72 & 43.74       \\
 \theiasecondbox{}\,\theiaEfive{}       & 388,181 & 0.04 & 160.8 & 2 & 1.24     \\
 \clearscopesecondbox{}\,\ClearscopeEfive{}  & 24,426 & 0.48 & 117.4  & 33 & 28.11    \\
\end{tblr}

\end{table}

\subsection{Detailed Investigation}\label{subsec:detailed_investigation}
Since the results for Theia E5 in the \attackrecall{} show that the attack was never reported, though our \gt{} investigation, we are examining this in more detail. 
In this way, we also demonstrate the benefits of our time-based alarms. 

\begin{figure}[h]
    \centering
    \includegraphics[width=.5\textwidth]{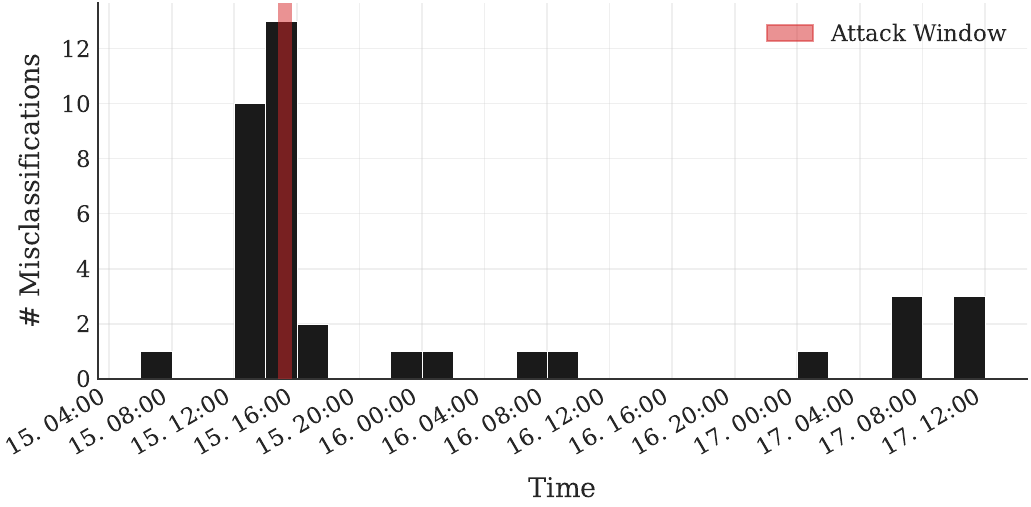}
    \caption{\grasp{}'s time reporting (default configuration).}
    \label{fig:theia_report}
 \end{figure}

Figure~\ref{fig:theia_report} shows a Theia E5 run in which no attack was detected after the \gt{} investigation. 
37 alarms were generated for 388,181 investigated processes, resulting in 47 time-based alarms, as some processes were reported in multiple time windows. 
Only two alarms are from unseen executables. 
A peak is visible on the 15th between 2\,p.m. and 4\,p.m.

 \begin{figure}[t]
    \centering
    \includegraphics[width=.48\textwidth]{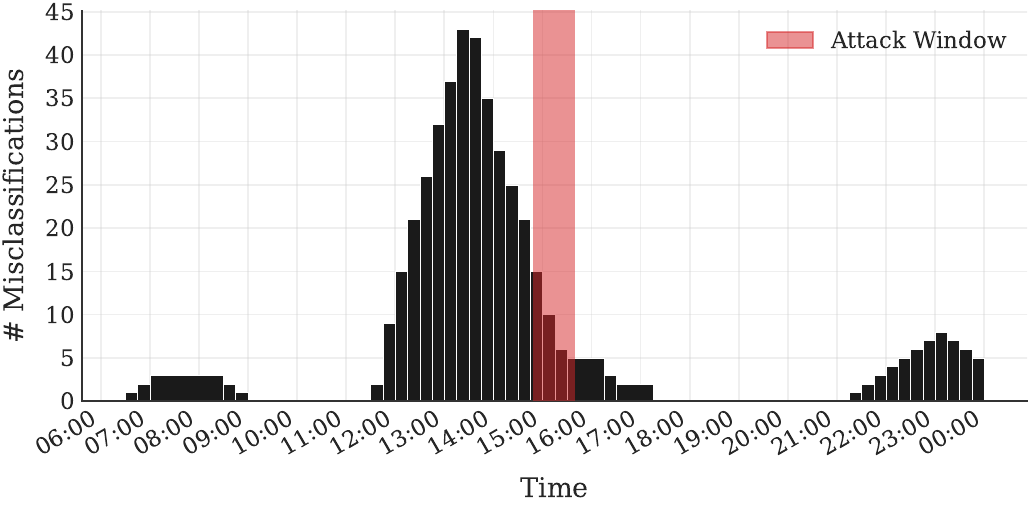}    
    \caption{\grasp{}'s time reporting (15-minute step size).}
    \label{fig:theia_report_detailed}
 \end{figure}

 We retrained the training data with finer granularity, using a context size of 120 and a 15-minute step size, and tested it only at the peak on day 15. 
 A total of 16 unique alarms were generated. 
 The corresponding alarms occurred multiple times in the time windows, resulting in 60 time-based alarms. These produce the following Figure~\ref{fig:theia_report_detailed}. 
 We see that the peak occurs slightly before the actual attack, but the attack is still detected. There may be several reasons for this: the \gt{} documentation may have been imprecise, the time periods may have been poorly documented, or preparatory measures of the attack may have been undocumented. 

 If we take a closer look at the alarms, we find clear indications of the attack, such as the misuse of \texttt{firefox}, which are also documented.  
 This was not reflected in our \gt{} investigation with time-based filtering, but the attack went through \texttt{firefox}, and four nodes with \texttt{firefox} as executable behaved too differently, so it was misclassified. 
 A more detailed investigation of the alarms is the goal of our future work, in which we plan to expand triage, backtracking, and visualization based on them. 

 \subsection{Hyperparameters Study}\label{subsec:hyperparamter}
Our hyperparameter study covers relevant adjustments for \grasp{}. These include neighborhood consideration, batch size, training epochs, context size, and step size.

\noindent{{\textbf{Neighborhoods:}}}
To examine the neighborhoods, in addition to the 3-hop neighborhood shown in Appendix~\ref{app:neighborhoods} and discussed in Section~\ref{sec:background}, we also considered a sampled variant of the two-hop neighborhood with 20 1-hop neighbors and up to 20 neighbors of those neighbors, and a sampled 1-hop neighborhood with 10,000. 

The results in Appendix~\ref{app:hyper}, Table~\ref{tab:hyper_studie}, indicate that increasing the number of neighbors leads to longer computation times without substantially improving detection quality. While some individual experiments show better results.
A sampled 2-hop neighborhood is considerably more computationally efficient and produces only slightly worse results, demonstrating strong potential for practical, efficient application. Furthermore, considering a one-hop neighborhood in general yields competitive performance across datasets, enabling the method to operate with significantly higher efficiency. We plan to investigate this in future work.

\noindent{{\textbf{Batch sizes:}}} 
We considered 8 and 512 in addition to our default of 32. There was no significant improvement in detection with a batch size of 8. With a batch size of 512, the learning of executables in particular deteriorated significantly, even though detection remained similarly good. These parameters can be adjusted to affect training time, and our default value of 32 represents a good balance between training time and detection performance in our experiments. 

\noindent{{\textbf{Training Epochs:}}}
The default is four training epochs; in the hyperparameter study, we examine both 2 and 8. 
We find no significant deterioration with shorter training, and a slight improvement with longer training. 
This is mainly because the training loss changes little after 2 epochs. 
Furthermore, we have a notably better detection performance on the Cadets~E3 dataset.
Overall, we see potential in trying different training strategies, especially for addressing class imbalance. 

\noindent{{\textbf{Context Size:}}} 
The context size specifies the maximum amount of neighborhood information available for classification; neighborhood size is used to sample and reduce it. Our default is 120 minutes, and we have also tried 60 and 15 minutes. 
As in our default, we set the step size equal to the context size. The results show that a smaller context size can reduce the neighborhood, but can also impede the learning of executable behavior. We attribute this to the fact that, depending on the selected context size, the focus is either on a shorter or a longer history of processes. In practice, an ensemble of different context sizes is interesting for detailed detection, which we plan to examine further in future work. 

\noindent{{\textbf{Step Size:}}} 
The step size of the sliding windows primarily determines detection granularity. This involves a trade-off between longer preprocessing and training. Not only does the training take longer, but it also sees more training data, as many processes are made available to the model more often. By default, the sliding window is not overlapping because we set it to the same context size of 120 minutes. In the experiments, we investigated a step size of 10 minutes for both a context size of 120 and 60 minutes. 

The results in Table~\ref{tab:hyper_studie} show that no significant improvement could be achieved. Our \MacroFone{} value indicates that it performs similarly to a step size equal to the context size. 

\subsection{Ablation Study}\label{subsec:hyperparameter}
In our ablation study, we deactivated system components one at a time to measure their effects. 
First, the transformer-based autoencoder for encoding location information. 
Second, the neighborhood, and finally, our clustering. 

We have presented the results in Table~\ref{tab:ablation_study}. 
It is interesting that the autoencoder also has a negative effect: the number of anomalies has decreased without it. 
We attribute this to the fact that changed bahavior, like communication with new file paths or IP addresses, isnt represented anymore, so it isnt detactable. 
This leads us to further investigate and implement a Word2Vec-based encoding, yielding results nearly identical to the transformer-based autoencoder but with a higher \MacroFone{} score, indicating it is better for our use case: learning the normal behavior of executables. 
Nevertheless, more alarms are generated compared to the untrained transformer. We attribute this to the fact that the lower information content only flags larger deviations from normal behavior.

As soon as we set the neighborhood to 1, i.e., consider only the process itself, we learned almost no process behavior, particularly for the \MacroFone{} metric. 
Nevertheless, we detected nearly all attacks, which can be attributed to conspicuous, unseen executables. 
This leads us to introduce a baseline that remembers all executables from the training data and reports all unseen ones as anomalies; the results are in Appendix~\ref{app:unseen_executables} in Table~\ref{tab:unseen_executables}.
Removing clustering from the architecture did not affect training; we were still able to learn the executables, but we generated more alarms and hits.

\begin{table}[h]
  \centering
  \caption{Ablation study results on \cadetsEthree{} and \theiaEthree{}. \label{tab:ablation_study}}
  
  \begin{tblr}{
    width=\linewidth,
    colspec = {l*{6}{r}},
    row{1} = {valign=m,c,font=\bfseries\scriptsize},
    cells={font=\scriptsize},
    hlines = {0.2pt, dashed},
    vlines = {0.2pt, dashed},
    hline{1,2,Z} = {0.5pt, solid},
    vline{1,2,Z} = {0.5pt, solid},
    hline{3,5,6,8,9,11,12}   = {0.5pt, solid},
    rowsep=1pt,
  }
    
  Dataset & \meanattackrecall{}  & \meanalarms{} & \cvalarms{} & \meanh{} & \meanFone{} & \meanMacroFone{} \\
  \SetCell[c=7]{c} Transformer-Based Autoencoder\\
  \cadetsfirstbox{}\,\cadetsEthree{}    &    1.00   &  143.0   &  0.39   & 18.8    &  0.98  & 0.43 \\
  \theiafirstbox{}{}\,\theiaEthree{}    &    1.00   &  192.8   &  0.10   & 4.0   &  0.94  & 0.20 \\
  \SetCell[c=7]{c} Neighborhood \\
  \cadetsfirstbox{}\,\cadetsEthree{}    &    1.00   &  143.0   &  0.88   & 12.0   &  0.16  & 0.01 \\
  \theiafirstbox{}{}\,\theiaEthree{}    &    1.00   &  388.0   &  0.40   & 4.0   &  0.42  & 0.02 \\
  \SetCell[c=7]{c} Clustering\\
  \cadetsfirstbox{}\,\cadetsEthree{}    &    1.00   &  1,663.8   &  0.09   & 26.0   &  0.98  & 0.43 \\
  \theiafirstbox{}{}\,\theiaEthree{}    &    1.00   &  14,819.2   &  0.28   & 21.2   &  0.96 & 0.22 \\
  \SetCell[c=7]{c} Word2Vec instead of Transformer-Based Autoencoder\\
  \cadetsfirstbox{}\,\cadetsEthree{}    &    1.00   &  477.2   &  0.09   & 23.8    &  0.99  & 0.49 \\
  \theiafirstbox{}{}\,\theiaEthree{}    &    1.00   &  343.0   &  0.10   & 71.4   &  0.95  & 0.22 \\
  \end{tblr}
  
\end{table}

\subsection{Living-off-the-Land Attack Study}
Since our system detects anomalies based on unseen executables, and most attacks in the evaluation datasets also use them, we conducted a separate evaluation to examine a LOTL attack. 
Unseen executables and their interactions were deleted, as it is hard to meaningfully mimic or integrate them into the graph structure. 
This approach is especially hard to detect because complete steps were removed and not replaced. 

The results in Table~\ref{tab:lotl} show that \grasp{} still functions well, the counts are lower, which relies mostly on the \gt{}, because most entries are based on unseen executables. The results are from three runs. 

\begin{table}[t]
  \centering
  \caption{\grasp{}'s on the LOTL modified datasets.\label{tab:lotl}}
  \begin{tblr}{
    width=\linewidth,
    colspec={|X[2.0,l]|*{4}{X[c]}|},
    row{1}={valign=m,c,font=\bfseries},
    cells={font=\scriptsize},
    hlines = {0.2pt, dashed},
    vlines = {0.2pt, dashed},
    hline{1,2,Z} = {0.5pt, solid},
    vline{1,2,Z} = {0.5pt, solid},
    rowsep=1pt,
  }
  Dataset & \meanattackrecall{} & \cvattackrecall{} & \meanalarms{} & \meanh{} \\
  \cadetsfirstbox{}\,\cadetsEthree{}{}     &    0.78   & 0.32     & 470.6  & 6.3 \\
  
  \theiafirstbox{}\,\theiaEthree{}{}     & 0.67 & 0.70 & 146.7  &  1.3 \\
  \clearscopefirstbox{}\,\ClearscopeEthree{}{}     & 0.00 & 0.00 & 106.0  & 0.0 \\
  
  \optcfirstbox{}\,\optcfirst{}{}       & 0.00 & 0.00 & 8.3  & 0.0 \\
  \optcsecondbox{}\,\optcsecond{}{}     & 1.00 & 0.00 & 150.7  & 3.7 \\
  \optcthirdbox{}\,\optcthrid{}{}       & 1.00 & 0.00 & 100.3  & 14.7 \\

  \cadetssecondbox{}\,\cadetsEfive{}     &    0.00   & 0.00     & 77.7   & 0.0 \\
  
  \theiasecondbox{}\,\theiaEfive{}     & 0.00 & 0.00 & 79.3  &  0.0 \\
  \clearscopesecondbox{}\,\ClearscopeEfive{}     & 0.75 & 0.00 & 121.3  & 24.3 \\
  \end{tblr}
  
\end{table}

\section{Discussion}\label{sec:discussion}
We discuss our findings regarding our shortcomings and address findings from the datasets and the LOTL study. 

\noindent{\textbf{SC\,1-SC\,4:}}
The first four SC are discussed in Subsection~\ref{subsec:sc_addressed}. SC\,1 (Insufficient Detection Granularity) is mitigated by our node-based, self-supervised classification approach with a configurable temporal resolution, which enables higher granularity than existing systems. 
SC\,2 (Impractical Thresholding Methods) is resolved by eliminating dependence on threshold determination, ensuring robustness and consistency across datasets. 
For SC\,3 (Featurization Methods Trained on Test Data), our preprocessing and transformer-based encoding strategies prevent data leakage while maintaining adaptability to unseen node and edge types. 
Finally, SC\,4 (Missing Anomaly Detection Capacity) is addressed by demonstrating improved anomaly coverage that captures conspicuous events that prior systems missed.

 \noindent{\textbf{SC\,5 Not Measuring Instability:}}
Instability was measured by performing five runs in the main investigation. In addition, important metrics, such as \attackrecall{}, were reported along with their corresponding coefficient of variation. 
Furthermore, we conducted additional investigations with three runs each.

\noindent{{\textbf{SC\,6 Missing Analysis of False Positives:}}}
\grasp{}, like related work, is an anomaly detection method, as in related work. 
We have also investigated its anomaly detection capability and achieved better results than related work, which ignores most anomalous reports.
The analysis of false alarms includes unknown executables, the number of alarms, and their influence on the total number of alarms.  
Future work will examine neighborhood sharing of executables across files, flows, and edges to improve alarm triage and backtracking.

\noindent{{\textbf{SC\,7 Selected Training and Validation Splits:}}}
For our scenario, we use all data prior to the first indication of an attack, and test the remaining data. 
We use the first day specified in the \gt{} for the respective dataset as the reference date. 
In practice, this could be reports from employees, firewalls, or other organizational entities. 
We do not consider it feasible to perform specific splits and arbitrarily select the validation dataset, on which the threshold may also depend. 

For the OpTC datasets, we had to use modified versions to compare them with related work, as their data processing requires all data to be stored in main memory. 
We tried to keep the scenario as realistic as possible and shortened the time periods in each case. 
We were unable to meaningfully shorten the fifth engagement, so we could not compare this scenario with related work on reevaluating the systems. The comparison based on the data from~\cite{velox}, in Table~\ref{tab:full_tblr}, is not affected by this. 

\noindent{\textbf{SC\,8 Reproducibility:}}
The reported results from prior work are not consistently reproducible when using robust metrics across multiple runs. 
The original authors themselves note that achieving those results depends on dataset-specific tuning and occasional lucky runs. 
While similar outcomes can sometimes be replicated, they are not reliable across runs. 
In contrast, \grasp{} provides robust reproducible results.

\noindent{\textbf{Datasets:}}
During the evaluation, we observed that \grasp{} cannot learn the executable behavior of the Clearscope datasets. 
This is due to the structure of the datasets: one executable strongly dominates each batch, while the others occur rarely, and the neighborhoods are hardly distinguishable. 
The OpTC datasets are difficult to learn, particularly because the end users are simulated, and their behavior is harder to predict. 
The diversity of executables and more random occurrences make learning and, thus, detection more difficult. 
The low \MacroFone{} values are mainly due to the high executable count and that the majority occur rarely, if not only once, in the training data. 
However, this is successfully covered by our clustering component, which bundles the behavior of rare and similar executables.
Furthermore, the attacks often rely on unknown executables, which motivates the creation of additional baseline datasets.

\noindent{\textbf{LOTL:}}
Since \grasp{} is designed to report all unknown executables, we examined the LOTL attack scenario, as particular APT actors also adapt to detection methods, so it cannot be assumed that the attacker will use novel executables. 
We removed unknown executables and paths and still achieved good detection. 
We can still detect attacks and even unusual behavior.
We attribute the fact that some attacks could no longer be detected to the removal of all processes containing unknown executables, which eliminated all suspicious activity.

\section{Conclusion}\label{sec:conclusion}
This paper introduced \grasp{}, a novel graph-based anomaly detection method that uses masked self-supervised classification of executables. 
By predicting executables from the neighborhood context, \grasp{} models system behavior without thresholds, effectively detecting attacks and anomalous process behavior.
While the method produces more alarms than related work, it achieves higher overall detection coverage and is more stable across different dataset splits. 
Moreover, we demonstrated that many of the alarms we delivered provided valuable insights into changes in system behavior, as they indicated deviations from the system’s prior normal behavior. 

Our results show that we can detect attacks using attack recall, except for the Clearscope dataset, where we cannot learn executable behavior from the Android dataset due to its structure. We also showed that, even when we do not detect the attack based on \gt{} evaluation, we can still report the attack effectively through a more detailed analysis. 

Future work should evaluate the method using additional datasets to test its generalizability beyond the DARPA scenarios. 
The LOTL attack study can also be expanded. Also, the training with class-imbalanced data. 
In addition, security analysts should evaluate the practical usefulness of the generated alarms in real-world threat hunting workflows.
We see particular potential in further analyzing the alarms to provide an explanation for them. We also see potential in triage, backtracking, and graph visualization. 
The combination of different context sizes for longer and shorter process histories offers a promising starting point for reducing the number of alarms through majority decision.
Continuous training strategies can further reduce false alarms by adapting to evolving system behavior. 

\section*{Ethical Considerations}

There are no ethical concerns from our perspective. 
All experiments are based on publicly available, ethically collected datasets that we use to evaluate intrusion detection systems and that do not contain any sensitive information. 
The provided system functions solely as a detection tool and likewise raises no ethical issues.

\section*{Open Science}
We release \grasp{} as open source under the MIT License to encourage community use and further development. 
The complete codebase and instructions for reproducing the experiments presented in this paper are available in our Git repository~\url{https://github.com/grbaande/grasp}.

\section*{LLM Usage Consideration}
During the preparation of this work, we used DeepL\footnote{\url{www.deepl.com}}, Grammarly\footnote{\url{www.grammarly.com}}, and Perplexity\footnote{\url{www.perplexity.ai}} to translate and rephrase text. After using these tools, we reviewed and edited the content as needed and take full responsibility for the content.

\bibliographystyle{IEEEtran}
\bibliography{./lib.bib}

\balance

\newpage

\appendices
\section{Neighborhoods}\label{app:neighborhoods}
A higher neighborhood consideration increases the amount of available information for classification. Within GNNs, oversmoothing occurs when neighborhoods become too large. Table~\ref{tab:neighborhoods} shows how neighborhoods behave as neighborhood determination increases. It should be noted that a higher neighborhood requires more system resources for determination. The system resources required for Cadets E5 were too large for our system, so we were unable to determine the 3-hop and 4-hop neighborhoods within a two-hour context window in a reasonable amount of time. We terminated the determination after 24 hours in each case.

\begin{table}[b!]
\centering

\caption{Statistics for available context information.\label{tab:neighborhoods}}
\begin{tblr}{
  width=\linewidth,
  colspec = {X[l]*{8}{r}},
  cells={font=\tiny},
  row{1,2} = {valign=m,c,font=\bfseries\tiny},
  hlines = {0.2pt, dashed},
  vlines = {0.2pt, dashed},
  hline{1,3,Z} = {0.5pt, solid},
  vline{1,2,Z} = {0.5pt, solid},
  vline{4,6,8} = {0.5pt, solid},
  colsep=4pt,
  rowsep=0pt,
}
\SetCell[r=2]{m}{Dataset}& \SetCell[c=2]{c}{1-hop} & & \SetCell[c=2]{c}{2-hop} & & \SetCell[c=2]{c}{3-hop} & & \SetCell[c=2]{c}{4-hop} & \\
 & Max & Mean & Max & Mean & Max & Mean & Max & Mean \\
 \cadetsfirstbox{}\,\cadetsEthree{}     & 33,749 & 16    & 35,367 & 1,567  & 65,892 & 6,555  & 65,921 & 6,612  \\
 \theiafirstbox{}\,\theiaEthree{}       & 142,133 & 20    & 202,755 & 2,077  & 208,863 & 13,505  & 215,303 & 31,702  \\
 \clearscopefirstbox{}\,\ClearscopeEthree{} &  9,502 & 39    & 9,515 & 764   & 12,516 & 6,861   & 12,516 & 6,889   \\
  \optcfirstbox{}\,\optcfirst{}{}      & 48,983 & 78    & 76,817 & 1,392  & 81,321 & 15,557  & 81,473 & 42,653  \\
 \optcsecondbox{}\,\optcsecond{}{}       & 48,072 & 66    & 61,304 & 1,318  & 68,178 & 16,493  & 69,272 & 43,291  \\
 \optcthirdbox{}\,\optcthrid{}{}       & 48,833 & 64    & 75,906 & 1,340  & 80,965 & 16,742  & 80,979 & 43,241  \\
 \cadetssecondbox{}\,\cadetsEfive{}     & 62,297 & 19    & 111,743 & 59,554    & n/a & n/a    & n/a & n/a    \\
 \theiasecondbox{}\,\theiaEfive{}      & 156,054 & 28    & 165,565 & 8,421    & 211,802 & 73,753    & 212,657 & 88,664    \\
 \clearscopesecondbox{}\,\ClearscopeEfive{} & 7,076 & 96    & 8,870 & 560    & 26,593 & 6237    & 26,604 & 6,941    \\
\end{tblr}

\end{table}

\section{Datasets}\label{app:dataset_stats}
In Table~\ref{tab:dataset_statistics}, we present statistics for the datasets we use. We look at the number of events before and after the filtering for events relevant to our analysts. We provide information on the number of processes, files, netflows, and possible executables. 
The datasets vary in the number of events.
The number of executables, our learning objective, is similar across the datasets, except for \ClearscopeEthree{}, which has 44 executables. 
There has been a change in the engagements, with those from the fifth being larger than those from the third.
The number of subjects is important to us, as it indicates how many batches we process and for which we determine the executable. 
The files and netflows serve as additional information for our classification. 

\begin{table}[b!]
  
  \caption{Dataset statistics.\label{tab:dataset_statistics}}
  \begin{tblr}{
    cells={font=\tiny},
    width=\linewidth,
    colspec = {lr*{5}{X[r]}},
    row{1} = {valign=m,c,font=\bfseries\tiny},
    hlines = {0.2pt, dashed},
    vlines = {0.2pt, dashed},
    hline{1,2,Z} = {0.5pt, solid},
    vline{1,2,Z} = {0.5pt, solid},
    rowsep=0pt,
    }
        Dataset & \# Raw Events & \# \velox{} Events & \# Subjects & \# Files & \# Netflows & {\#~Executables}  \\
        \cadetsfirstbox{}\,\cadetsEthree{}  & 36,480,130   & 18,699,410   & 224,146  & 2,303,164 & 155,322  & 130 \\
        
        \theiafirstbox{}\,\theiaEthree{}   & 44,373,491   & 39,884,746   & 278,363  & 793,899   & 186,100  & 168 \\
        
        \clearscopefirstbox{}\,\ClearscopeEthree{}  & 18,317,851   & 18,316,129   & 12,469   & 76,938    & 279,694  & 44 \\
        
        \optcfirstbox{}\,\optcfirst{} & 31,609,342   & 31,611,029   & 50,626   & 380,963   & 4,184,346 & 201 \\
        \optcsecondbox{}\,\optcsecond{} & 38,133,406   & 38,151,059   & 99,693   & 472,501   & 4,696,258 & 168 \\
        \optcthirdbox{}\,\optcthrid{} & 28,050,992   & 28,049,214   & 70,913   & 345,365   & 3,547,821 & 186 \\
        \cadetssecondbox{}\,\cadetsEfive{}  & 1,075,396,080 & 751,395,597 & 6,576,457 & 8,434,050 & 219,384  & 126 \\
        \theiasecondbox{}\,\theiaEfive{}   & 140,994,301  & 139,736,402  & 1,264,440 & 984,937   & 36,747   & 118 \\
        \clearscopesecondbox{}\,\ClearscopeEfive{}  & 140,994,301  & 198,794,211  & 73,097   & 324,210   & 104,022  & 109 \\
    \end{tblr}
    
\end{table}

\section{Dataset Splits}\label{app:splits}
Since the dataset is split into training, validation, and test sets for evaluation, and these splits are handled differently in related work, we list the relevant splits for this work in Table~\ref{tab:dataset_splits}. 
We have divided our dataset into training and test sets only, as this is more practical. If a method relies on validation data, some of the training data is used for validation.

\begin{table}[h!]
  \caption{Dataset splits and attack count.\label{tab:dataset_splits}}
  \tiny
  \begin{tblr}{
    width=\linewidth,
    colspec={l*{8}{X[r]}},
    cells = {valign=m,font=\scriptsize},
    row{1}={valign=m,c,font=\bfseries\tiny},
    hlines = {0.2pt, dashed},
    vlines = {0.2pt, dashed},
    hline{1,2,Z} = {0.5pt, solid},
    vline{1,2,Z} = {0.5pt, solid},
    rowsep=1pt,
    colsep=1pt,
    }
        Dataset & {\#~At-\\tacks} & \velox{} days train & \velox{} days val & \velox{} days test & \grasp{} days train  & \grasp{} days test   & Modified \grasp{} days train &  Modified \grasp{} days test   \\
        \cadetsfirstbox{}\,\cadetsEthree{}  & 4   & 02-05, 07-09    & 10  & {06,\\11-13} & 02-05  & 06-13 & n/a  & n/a \\
        \theiafirstbox{}\,\theiaEthree{}   & 2   & 03-08   & 09  & {10, 12,\\13}   & 03-09  & 10-13 & n/a  & n/a \\
        \clearscopefirstbox{}\,\ClearscopeEthree{}  & 2   & 03-05, 07-10   & 02   & 11, 12    & 2-10  & 11-13 & n/a  & n/a \\
        \optcfirstbox{}\,\optcfirst{} & 1   & 19-21   & 22   & 23-25   & 16-24 & 25 & 22-24 & 25 \\
        \optcsecondbox{}\,\optcsecond{} & 1   & 19-21   & 22   & 23-25   & 16-22 & 23-25 & 21-22 & 23-25 \\
        \optcthirdbox{}\,\optcthrid{} & 1   & 19-21   & 22   & 23-25   & 18-23 & 24-25 & 22-23 & 24-25 \\
        \cadetssecondbox{}\,\cadetsEfive{}  & 2 & {08, 09,\\11} & 12 & 16, 17 & 07-14  & 15-17 & n/a & n/a \\
        \theiasecondbox{}\,\theiaEfive{}   & 1  & 08-10  & 11 & 14, 15   & 07-14   & 15-17 & n/a   & n/a \\
        \clearscopesecondbox{}\,\ClearscopeEfive{}  & 3  & 08-12  & 13   & {14, 15,\\17}   & 07-14  & 15-20 & n/a  & n/a \\
    \end{tblr}
\end{table}

\section{Used Components of the Provenance Data}\label{app:graph_info}
We have adopted the component pre-filtering of \orthrus{} and \velox{}. We use the nodes Subject, Netflows, and Files. For the datasets of the third and fifth engagements, we use the edge types: 
connect, execute, open, read, recvfrom, recvmsg, sendmsg, sendto, write, clone, and, for the OpTC datasets, the edge types:
open, read, create, message, modify, start, rename, delete, terminate, and write. 
The features used are either the command for an executable or its path. 
For the OpTC and Theia datasets, the path is used because it specifies the executable; for Cadets and Clearscope, the cmd is used because it specifies the executable in those cases. For netflows, we use the destination IP and port. The files have the corresponding path information. 

\section{Deep Learning Methodology}\label{app:dl_methodology}

\textbf{Topology}:
Graph Attention Network (GAT) as encoder and a Multi-Layer Perceptron (MLP) decoder for classification.
\begin{itemize}
  \item GAT Layer 0: $d_{in} \to 128$ channels $\times$ 4 heads (512 concatenated)
  \item GAT Layer 1: 512 $\to$ 128 $\times$ 4 heads (512 concatenated)
  \item MLP Head: 512 $\to$ 128 $\to$ $K$ (softmax)
  \item After each Layer: ReLU activation
  \item Dropout: 0.1 (only for GAT)
\end{itemize}
Cross-Entropy loss for multi-class classification (known executables) was used.

\textbf{Input Features}:
Operation type (1-hot) as Edge Features, node type (1-hot), location embeddings (8-D from pre-trained transformer) and Executable (1-hot in a masked setting, so target node never knows his executable).
Edges were treated as undirected.

\textbf{Pre-training}: Location Transformer autoencoder (Transformer encoder-decoder, 8-D embeddings, 2 layers, 4 heads, 10 epochs) pre-trained on character-level location sequences (max 100 chars) with reconstruction loss.

\textbf{Optimizer}: Adam, $\eta=0.01$, weight decay $=10^{-4}$.

\textbf{Scheduler}: ReduceLROnPlateau, factor=0.5, patience=5 epochs.

\textbf{Batching}: 
Batch size: Neighborhood with nodes and edges based on n input nodes (default, 32);
Method: PyTorch Geometric NeighborLoader (default all neighbors, [-1, -1]);
Input nodes: process nodes only.

\textbf{Training Data Ordering}: Window-based (default 120-minute context, 120-minute stride). We shuffle the batches while training. 

\textbf{Stopping Criterion}: Fixed 4 epochs (default).

\textbf{Clustering after Training}: Misclassification, based on inference on the training data where stored in a dictonary, where each executable has a list of misclassified nodes. 

\textbf{Inference}
Misclassification on the test data where treated as anomalies. A misclassification occurs when the predicted executable does not match the true executable and the missclassification also is not included in the clustering as benign mixup.

\textbf{Implementation}
We used PyTorch (2.8.0) with PyTorch Geometric (2.7.0), together with CUDA-specific PyG extension packages compiled for PyTorch 2.8/CUDA 12.9: torch\_cluster (1.6.3+pt28cu129), torch\_scatter (2.1.2+pt28cu129), torch\_sparse (0.6.18+pt28cu129), and torch\_spline\_conv (1.2.2+pt28cu129).

\section{Baseline on Unseen Executables}\label{app:unseen_executables}
A simple baseline approach could involve flagging all new executables, which are not included in the training dataset. 
The results are shown in Table~\ref{tab:unseen_executables}. 
Since this is a static method, the results are absolute and not averaged across multiple runs. 
OpTC, Cadets E5, and Theia E3 deliver good results in this baseline. 
This is primarily due to the attacks carried out and the \gt{}. 
However, since we cannot rely on the attacker using novel executable files, this merely serves as an incentive to create new datasets. 
Furthermore, our \grasp{} results are close to the baseline and yield more anomalies and more true positives. 
However, since new executables also communicate with existing ones, this can lead to behavioral differences. 

\begin{table}[b!]
\centering
\caption{Report unseen executables evaluation.\label{tab:unseen_executables}}
  \begin{tblr}{
    width=\linewidth,
    colspec={|X[2.0,l]|*{3}{X[c]}|},
    row{1}={valign=m,c,font=\bfseries},
    cells={font=\scriptsize},
    hlines = {0.2pt, dashed},
    vlines = {0.2pt, dashed},
    hline{1,2,Z} = {0.5pt, solid},
    vline{1,2,Z} = {0.5pt, solid},
    rowsep=0pt,
  }
  Dataset & \ensuremath{\mathrm{A_{r}}} &  {\ensuremath{a}} & \ensuremath{TP}\\
  \cadetsfirstbox{}\,\cadetsEthree{}{}     &    1.00       & 71  & 15\\
  
  \theiafirstbox{}\,\theiaEthree{}{}     & 1.00 & 150  &  69\\
  \clearscopefirstbox{}\,\ClearscopeEthree{}{}     & 0.00  & 1557  & 0 \\
  
  \optcfirstbox{}\,\optcfirst{}{}       & 1.00 &  106  & 36 \\
  \optcsecondbox{}\,\optcsecond{}{}     & 1.00 &  273  & 2 \\
  \optcthirdbox{}\,\optcthrid{}{}       & 1.00 &  86  & 17 \\

  \cadetssecondbox{}\,\cadetsEfive{}     &    1.00       & 72   & 15\\
  
  \theiasecondbox{}\,\theiaEfive{}     & 0.00 & 2  &  0\\
  \clearscopesecondbox{}\,\ClearscopeEfive{}     & 0.50  & 33  & 9 \\
  \end{tblr}
  
\end{table}

\section{Default Configuration}\label{app:parameters}
Our default configuration uses a two-hour context size with a two-hour step size, in a non-overlapping setup. The batch size is 32, we trained for four epochs, and our neighborhood sampling strategy is tailored to the datasets, as large datasets cannot be efficiently processed with a complete two-hop neighborhood. Therefore, Cadets E3, Clearscope E3, and the OpTC data sets have a complete neighborhood view. The other data sets, Theia E3, Cadets, Clearscope, and Theia E5, are analyzed with a maximum of 10,000 1-hop neighbors and 20 2-hop neighbors each. 

\clearpage
\section{\velox{} Split Comparison}\label{app:velox}
\begin{table}[t!]
  \centering
 
  \caption{Detailed detection performance metrics evaluated with the \velox{} split. All values, except those from \grasp{}, are from Bilot et al.~\cite{velox}.}
  \label{tab:full_tblr}
  \tiny
  \begin{tblr}{
    width=\linewidth,
    colspec={|X[2.5,l]|X[c]X[c]X[c]|r|X[c]X[c]X[c]|X[r]X[r]|},
    row{1,2}={valign=m,c,font=\bfseries\tiny},
    rowsep=0pt,
  }
    \hline
    \SetCell[r=2]{m} System &
    \SetCell[c=3]{c}{ADP} & & & 
    \SetCell[r=2]{m}{\ensuremath{\delta}} &
    \SetCell[c=3]{c}{Precision} & & & 
    \SetCell[c=1]{l}{TP}&
    \SetCell[c=1]{l}{FP} \\

    \SetCell[c=1]{l}{} &
    \SetCell[c=1]{c}{Mean} & 
    \SetCell[c=1]{c}{Min} & 
    \SetCell[c=1]{c}{Best} & 
    \SetCell[c=1]{l}{}&
    \SetCell[c=1]{c}{Mean} & 
    \SetCell[c=1]{c}{Min} & 
    \SetCell[c=1]{c}{Best} & 
    \SetCell[c=2]{c}{Best} \\
    \hline
    \SetRow{abovesep=1pt}\SetCell[c=10]{c}\cadetsfirstbox{}\,\cadetsEthree{} \\
    \hline
    \SIGL{}         & \gc{0.00}0.00 & \gc{0.00}0.00 & \gc{0.01}0.01 & 8\%   & \gc{0.00}0.00 & \gc{0.00}0.00 & \gc{0.00}0.00 & 0    & 45    \\
    \threatrace{}  & \gc{0.01}0.01 & \gc{0.01}0.01 & \gc{0.01}0.01 & 36\%  & \gc{0.00}0.00 & \gc{0.00}0.00 & \gc{0.00}0.00 & 26   & 17k   \\
    \nodlink{}      & \gc{0.33}0.33 & \gc{0.15}0.15 & \gc{0.96}0.96 & 56\%  & \gc{0.00}0.00 & \gc{0.00}0.00 & \gc{0.00}0.00 & 50   & 55k   \\
    \magic{}        & \gc{0.03}0.03 & \gc{0.02}0.02 & \gc{0.06}0.06 & 37\%  & \gc{0.00}0.00 & \gc{0.00}0.00 & \gc{0.00}0.00 & 68   & 144k  \\
    \kairos{}       & \gc{0.01}0.01 & \gc{0.01}0.01 & \gc{0.01}0.01 & 24\%  & \gc{0.00}0.00 & \gc{0.00}0.00 & \gc{0.00}0.00 & 0    & 8     \\
    \flash{}        & \gc{0.10}0.10 & \gc{0.00}0.00 & \gc{0.34}0.34 & 127\% & \gc{0.05}0.05 & \gc{0.00}0.00 & \gc{0.25}0.25 & 1    & 3     \\
    \rcaid{}       & \gc{0.21}0.21 & \gc{0.04}0.04 & \gc{0.44}0.44 & 64\%  & \gc{0.21}0.21 & \gc{0.00}0.00 & \gc{1.00}1.00 & 2    & 0     \\
    \orthrus{}      & \gc{0.94}0.94 & \gc{0.85}0.85 & \gc{1.00}1.00 & 12\%  & \gc{0.66}0.66 & \gc{0.40}0.40 & \gc{1.00}1.00 & 10   & 0     \\
    \velox{}        & \gc{0.94}0.94 & \gc{0.77}0.77 & \gc{1.00}1.00 &  9\%  & \gc{0.85}0.85 & \gc{0.44}0.44 & \gc{1.00}1.00 & 8    & 0     \\
    \textbf{\grasp{}}        & \gc{1.00}1.00 & \gc{1.00}1.00 & \gc{1.00}1.00 &  0\%  & \gc{0.08}0.08 & \gc{0.08}0.07 & \gc{0.11}0.11 & 19    & 160    \\
    \hline
       \SetRow{abovesep=1pt}\SetCell[c=10]{c} \theiafirstbox{}\,\theiaEthree{} \\
    \hline
    \SIGL{}          & \gc{0.30}0.30 & \gc{0.25}0.25 & \gc{0.50}0.50 & 33\%  & \gc{0.18}0.18 & \gc{0.10}0.10 & \gc{0.33}0.33 & 1    & 2     \\
    \threatrace{}  & \gc{0.03}0.03 & \gc{0.01}0.01 & \gc{0.10}0.10 & 143\%  & \gc{0.00}0.00 & \gc{0.00}0.00 & \gc{0.00}0.00 & 118   & 701k  \\
    \nodlink{}     & \gc{0.24}0.24 & \gc{0.00}0.00 & \gc{0.50}0.50 & 90\%  & \gc{0.00}0.00 & \gc{0.00}0.00 & \gc{0.00}0.00 & 24   & 199k  \\
    \magic{}        & \gc{0.00}0.00 & \gc{0.00}0.00 & \gc{0.00}0.00 &  0\%  & \gc{0.00}0.00 & \gc{0.00}0.00 & \gc{0.00}0.00 & 74   & 196k  \\
    \kairos{}       & \gc{0.45}0.45 & \gc{0.25}0.25 & \gc{0.50}0.50 & 22\%  & \gc{0.70}0.70 & \gc{0.50}0.50 & \gc{1.00}1.00 & 2    & 0     \\
    \flash{}        & \gc{0.02}0.02 & \gc{0.01}0.01 & \gc{0.05}0.05 & 99\%  & \gc{0.00}0.00 & \gc{0.00}0.00 & \gc{0.00}0.00 & 3    & 8.2k  \\
    \orthrus{}      & \gc{0.44}0.44 & \gc{0.10}0.10 & \gc{1.00}1.00 & 73\%  & \gc{0.24}0.24 & \gc{0.07}0.07 & \gc{1.00}1.00 & 8    & 0     \\
    \velox{}        & \gc{0.72}0.72 & \gc{0.42}0.42 & \gc{0.97}0.97 & 30\%  & \gc{0.53}0.53 & \gc{0.03}0.03 & \gc{0.91}0.91 & 10   & 1     \\
    \textbf{\grasp{}}       & \gc{1.00}1.00 & \gc{1.00}1.00 & \gc{1.00}1.00 &  0\%  & \gc{0.28}0.28 & \gc{0.22}0.22 & \gc{0.33}0.33 & 65    & 132     \\
    \hline
    \SetRow{abovesep=1pt}\SetCell[c=10]{c}\clearscopefirstbox{}\,\ClearscopeEthree{} \\
    \hline
    \SIGL{}          & \gc{0.02}0.02 & \gc{0.01}0.01 & \gc{0.02}0.02 & 6\%   & \gc{0.00}0.00 & \gc{0.00}0.00 & \gc{0.01}0.01 & 3    & 391   \\
    \threatrace{}  & \gc{0.02}0.02 & \gc{0.01}0.01 & \gc{0.03}0.03 & 56\%  & \gc{0.00}0.00 & \gc{0.00}0.00 & \gc{0.00}0.00 & 41   & 35k   \\
    \nodlink{}     & \gc{0.01}0.01 & \gc{0.00}0.00 & \gc{0.03}0.03 &121\%  & \gc{0.00}0.00 & \gc{0.00}0.00 & \gc{0.00}0.00 & 0    & 0     \\
    \magic{}      & \gc{0.02}0.02 & \gc{0.00}0.00 & \gc{0.03}0.03 & 52\%  & \gc{0.00}0.00 & \gc{0.00}0.00 & \gc{0.00}0.00 & 0    & 0     \\
    \kairos{}       & \gc{0.42}0.42 & \gc{0.04}0.04 & \gc{1.00}1.00 & 86\%  & \gc{0.00}0.00 & \gc{0.00}0.00 & \gc{0.00}0.00 & 0    & 0     \\
    \flash{}        & \gc{0.03}0.03 & \gc{0.01}0.01 & \gc{0.07}0.07 & 77\%  & \gc{0.00}0.00 & \gc{0.00}0.00 & \gc{0.00}0.00 & 35   & 23k   \\
    \rcaid{}       & \gc{0.14}0.14 & \gc{0.02}0.02 & \gc{0.50}0.50 &134\%  & \gc{0.03}0.03 & \gc{0.00}0.00 & \gc{0.08}0.08 & 15   & 169   \\
    \orthrus{}      & \gc{0.75}0.75 & \gc{0.25}0.25 & \gc{1.00}1.00 & 42\%  & \gc{0.00}0.00 & \gc{0.00}0.00 & \gc{0.00}0.00 & 1    & 399   \\
    \velox{}        & \gc{0.30}0.30 & \gc{0.02}0.02 & \gc{1.00}1.00 &116\%  & \gc{0.00}0.00 & \gc{0.00}0.00 & \gc{0.00}0.00 & 1    & 913   \\
    \textbf{\grasp{}}        & \gc{0.00}0.00 & \gc{0.00}0.00 & \gc{0.00}0.00 &  0\%  & \gc{0.00}0.00 & \gc{0.00}0.00 & \gc{0.00}0.00 & 0    & 0     \\
    \hline
    \SetRow{abovesep=1pt}\SetCell[c=10]{c}        \optcfirstbox{}\,\optcfirst{} \\
    \hline
    \threatrace{}         & \gc{0.57}0.57 & \gc{0.03}0.03 & \gc{1.00}1.00 & 66\%  & \gc{0.00}0.00 & \gc{0.00}0.00 & \gc{0.00}0.00 & 114  & 1.5M   \\
    \nodlink{}  & \gc{0.20}0.20 & \gc{0.00}0.00 & \gc{1.00}1.00 &197\%  & \gc{0.00}0.00 & \gc{0.00}0.00 & \gc{0.00}0.00 & 20   & 102k   \\
    \magic{}      & \gc{0.37}0.37 & \gc{0.20}0.20 & \gc{0.50}0.50 & 30\%  & \gc{0.00}0.00 & \gc{0.00}0.00 & \gc{0.00}0.00 & 102  & 259k   \\
    \kairos{}        & \gc{0.50}0.50 & \gc{0.11}0.11 & \gc{1.00}1.00 & 81\%  & \gc{0.14}0.14 & \gc{0.00}0.00 & \gc{0.33}0.33 & 1    & 2      \\
    \flash{}       & \gc{0.14}0.14 & \gc{0.03}0.03 & \gc{0.50}0.50 &130\%  & \gc{0.00}0.00 & \gc{0.00}0.00 & \gc{0.00}0.00 & 2    & 9.2k   \\
    \orthrus{}        & \gc{0.68}0.68 & \gc{0.06}0.06 & \gc{1.00}1.00 & 59\%  & \gc{0.08}0.08 & \gc{0.00}0.00 & \gc{0.20}0.20 & 1    & 4      \\
    \velox{}       & \gc{0.57}0.57 & \gc{0.11}0.11 & \gc{1.00}1.00 & 64\%  & \gc{0.12}0.12 & \gc{0.00}0.00 & \gc{0.50}0.50 & 1    & 1      \\
    \textbf{\grasp{}}       & \gc{1.00}1.00 & \gc{1.00}1.00 & \gc{1.00}1.00 &  0\%  & \gc{0.14}0.14 & \gc{0.11}0.11 & \gc{0.16}0.16 & 31    & 160     \\

    \hline
    \SetRow{abovesep=1pt}\SetCell[c=10]{c} \optcsecondbox{}\,\optcsecond{}\\
    \hline
    \threatrace{}         & \gc{0.13}0.13 & \gc{0.12}0.12 & \gc{0.17}0.17 & 12\%  & \gc{0.00}0.00 & \gc{0.00}0.00 & \gc{0.00}0.00 & 2.9k & 1.4M   \\
    \nodlink{}  & \gc{0.32}0.32 & \gc{0.08}0.08 & \gc{1.00}1.00 &108\%  & \gc{0.02}0.02 & \gc{0.02}0.02 & \gc{0.02}0.02 & 1.4k & 78k    \\
    \magic{}      & \gc{0.87}0.87 & \gc{0.33}0.33 & \gc{1.00}1.00 & 30\%  & \gc{0.00}0.00 & \gc{0.00}0.00 & \gc{0.00}0.00 & 1.1k & 1.2M   \\
    \kairos{}        & \gc{0.75}0.75 & \gc{0.25}0.25 & \gc{1.00}1.00 & 42\%  & \gc{0.38}0.38 & \gc{0.00}0.00 & \gc{1.00}1.00 & 1    & 0      \\
    \flash{}       & \gc{1.00}1.00 & \gc{1.00}1.00 & \gc{1.00}1.00 & 0\%  & \gc{0.00}0.00  & \gc{0.00}0.00 & \gc{0.01}0.01  & 1.5k & 239k     \\
    \orthrus{}        & \gc{0.48}0.48 & \gc{0.23}0.23 & \gc{1.00}1.00 & 57\%  & \gc{0.14}0.14 & \gc{0.04}0.04 & \gc{0.25}0.25 & 1 & 3   \\
    \velox{}       & \gc{0.90}0.90 & \gc{0.50}0.50 & \gc{1.00}1.00 &22\%  & \gc{0.18}0.18 & \gc{0.11}0.11 & \gc{0.25}0.25 & 1    & 3      \\
    \textbf{\grasp{}}        & \gc{1.00}1.00 & \gc{1.00}1.00 & \gc{1.00}1.00 &  0\%  & \gc{0.01}0.01 & \gc{0.01}0.01 & \gc{0.01}0.01 & 13    & 444     \\

    \hline
    \SetRow{abovesep=1pt}\SetCell[c=10]{c} \optcthirdbox{}\,\optcthrid{}\\
    \hline
    \threatrace{}         & \gc{0.62}0.62 & \gc{0.25}0.25 & \gc{1.00}1.00 & 52\%  & \gc{0.01}0.01 & \gc{0.00}0.00 & \gc{0.05}0.05 & 2    & 42     \\
    \nodlink{}  & \gc{0.03}0.03 & \gc{0.01}0.01 & \gc{0.08}0.08 & 82\%  & \gc{0.00}0.00 & \gc{0.00}0.00 & \gc{0.00}0.00 & 355  & 102k   \\
    \magic{}      & \gc{1.00}1.00 & \gc{1.00}1.00 & \gc{1.00}1.00 & 0\%  & \gc{0.00}0.00 & \gc{0.00}0.00 & \gc{0.00}0.00 & 185  & 260k   \\
    \kairos{}        & \gc{0.38}0.38 & \gc{0.09}0.09 & \gc{1.00}1.00 & 82\%  & \gc{0.12}0.12 & \gc{0.00}0.00 & \gc{0.33}0.33 & 1  & 2   \\
    \flash{}       & \gc{0.58}0.58 & \gc{0.14}0.14 & \gc{1.00}1.00 & 62\%   & \gc{0.00}0.00 & \gc{0.00}0.00 & \gc{0.00}0.00 & 222  & 215k      \\
    \orthrus{}        & \gc{0.25}0.25 & \gc{0.14}0.14 & \gc{0.33}0.33 & 29\%  & \gc{0.11}0.11 & \gc{0.04}0.04 & \gc{0.20}0.20 & 1    & 4   \\
    \velox{}       & \gc{0.39}0.39 & \gc{0.20}0.20 & \gc{0.50}0.50 & 34\%  & \gc{0.13}0.13 & \gc{0.00}0.00 & \gc{0.20}0.20 & 1    & 4      \\
    \textbf{\grasp{}}        & \gc{1.00}1.00 & \gc{1.00}1.00 & \gc{1.00}1.00 &  0\%  & \gc{0.03}0.03 & \gc{0.03}0.03 & \gc{0.05}0.05 & 10    & 212     \\
    \hline
    \SetRow{abovesep=1pt}\SetCell[c=10]{c} \cadetssecondbox{}\,\cadetsEfive{}\\
    \hline
    \threatrace{} & \gc{0.00}0.00 & \gc{0.00}0.00 & \gc{0.00}0.00 & 199\% & \gc{0.00}0.00 & \gc{0.00}0.00 & \gc{0.00}0.00 & 98 & 3.1M \\ 
    \nodlink{}     & \gc{0.23}0.23 & \gc{0.18}0.18 & \gc{0.33}0.33 & 24\% & \gc{0.00}0.00 & \gc{0.00}0.00 & \gc{0.00}0.00 & 76 & 708k \\ 
    \magic{}     & \gc{0.05}0.05 & \gc{0.03}0.03 & \gc{0.07}0.07 & 26\% & \gc{0.00}0.00 & \gc{0.00}0.00 & \gc{0.00}0.00 & 118 & 2.6M \\ 
    \kairos{}         & \gc{0.02}0.02 & \gc{0.01}0.01 & \gc{0.03}0.03 & 49\% & \gc{0.00}0.00 & \gc{0.00}0.00 & \gc{0.00}0.00 & 0 & 6 \\ 
    \flash{}       & \gc{0.04}0.04 & \gc{0.02}0.02 & \gc{0.04}0.04 & 21\% & \gc{0.00}0.00 & \gc{0.00}0.00 & \gc{0.00}0.00 & 6 & 24k \\ 
    \orthrus{}          & \gc{0.44}0.44 & \gc{0.26}0.26 & \gc{0.54}0.54 & 27\% & \gc{0.43}0.43 & \gc{0.25}0.25 & \gc{0.50}0.50 & 1 & 1 \\ 
    \velox{}            & \gc{0.31}0.31 & \gc{0.10}0.10 & \gc{0.50}0.50 & 51\% & \gc{0.12}0.12 & \gc{0.10}0.10 & \gc{0.14}0.14 & 11 & 68 \\ 
    \textbf{\grasp{}}        & \gc{1.0}1.0 & \gc{1.0}1.0 & \gc{1.0}1.0 &  0\%  & \gc{0.04}0.04 & \gc{0.07}0.04 & \gc{0.05}0.05 & 8    & 144     \\
    \hline   
    \SetRow{abovesep=1pt}\SetCell[c=10]{c} \theiasecondbox{}\,\theiaEfive{}\\
    \hline
    \threatrace{}             & \gc{0.05}0.05 & \gc{0.05}0.05 & \gc{0.05}0.05 & 2\% & \gc{0.00}0.00 & \gc{0.00}0.00 & \gc{0.00}0.00 & 64 & 722k \\
    \nodlink{}          & \gc{0.00}0.00 & \gc{0.00}0.00 & \gc{0.00}0.00 & 0\% & \gc{0.00}0.00 & \gc{0.00}0.00 & \gc{0.00}0.00 & 36 & 38k \\
    \magic{}         & \gc{0.20}0.20 & \gc{0.00}0.00 & \gc{1.00}1.00 & 199\% & \gc{0.00}0.00 & \gc{0.00}0.00 & \gc{0.00}0.00 & 68 & 445k \\
    \kairos{}          & \gc{0.17}0.17 & \gc{0.01}0.01 & \gc{0.50}0.50 & 112\% & \gc{0.03}0.03 & \gc{0.00}0.00 & \gc{0.08}0.08 & 2 & 23 \\
    \flash{}          & \gc{0.01}0.01 & \gc{0.01}0.01 & \gc{0.02}0.02 & 44\% & \gc{0.00}0.00 & \gc{0.00}0.00 & \gc{0.00}0.00 & 31 & 311k \\
    \orthrus{}          & \gc{0.70}0.70 & \gc{0.50}0.50 & \gc{1.00}1.00 & 31\% & \gc{0.39}0.39 & \gc{0.33}0.33 & \gc{0.50}0.50 & 1 & 1 \\
    \velox{}        & \gc{1.00}1.00 & \gc{1.00}1.00 & \gc{1.00}1.00 & 0\% & \gc{1.00}1.00 & \gc{1.00}1.00 & \gc{1.00}1.00 & 2 & 0 \\
    \textbf{\grasp{}}        & \gc{0.20}0.20 & \gc{0.00}0.00 & \gc{0.20}0.20 &  200\%  & \gc{0.00}0.00 & \gc{0.00}0.00 & \gc{0.00}0.01 & 2    & 294     \\
    \hline   
    \SetRow{abovesep=1pt}\SetCell[c=10]{c} \clearscopesecondbox{}\,\ClearscopeEfive{}\\
    \hline
    \threatrace{}             & \gc{0.01}0.01 & \gc{0.01}0.01 & \gc{0.01}0.01 & 19\% & \gc{0.00}0.00 & \gc{0.00}0.00 & \gc{0.00}0.00 & 42 & 64k \\
    \nodlink{}          & \gc{0.10}0.10 & \gc{0.01}0.01 & \gc{0.17}0.17 & 65\% & \gc{0.00}0.00 & \gc{0.00}0.00 & \gc{0.00}0.00 & 33 & 38k \\
    \magic{}       & \gc{0.01}0.01 & \gc{0.01}0.01 & \gc{0.01}0.01 & 11\% & \gc{0.00}0.00 & \gc{0.00}0.00 & \gc{0.00}0.00 & 50 & 105k \\
    \kairos{}          & \gc{0.33}0.33 & \gc{0.33}0.33 & \gc{0.33}0.33 & 0\% & \gc{0.12}0.12 & \gc{0.08}0.08 & \gc{0.25}0.25 & 1 & 3 \\
    \flash{}            & \gc{0.03}0.03 & \gc{0.03}0.03 & \gc{0.04}0.04 & 19\% & \gc{0.00}0.00 & \gc{0.00}0.00 & \gc{0.00}0.00 & 24 & 24k \\
    \orthrus{}         & \gc{0.09}0.09 & \gc{0.06}0.06 & \gc{0.12}0.12 & 25\% & \gc{0.04}0.04 & \gc{0.00}0.00 & \gc{0.10}0.10 & 1 & 9 \\
    \velox{}            & \gc{0.45}0.45 & \gc{0.25}0.25 & \gc{0.61}0.61 & 31\% & \gc{0.34}0.34 & \gc{0.21}0.21 & \gc{0.38}0.38 & 8 & 13 \\
    \textbf{\grasp{}}        & \gc{0.75}0.75 & \gc{0.75}0.75 & \gc{0.75}0.75 &  0\%  & \gc{0.18}0.18 & \gc{0.09}0.09 & \gc{0.34}0.34 & 35    & 67     \\
    \hline  
        
  \end{tblr}
   
\end{table}
To ensure fair comparability between previous related work and \grasp{}, we decided to apply our results to the dataset split used by Bilot et al.~\cite{velox}, using their \gt{} and metrics. 
Their work provides an overview comparison of previous PIDS, and the results of the individual systems are presented with optimized hyperparameters. We present the results of the overview study, with those from our \grasp{} system shown in Table~\ref{tab:full_tblr}. The \ensuremath{\delta} metric in Table~\ref{tab:full_tblr} stands for \cvADP{}\ensuremath{ \times 100}. 
The \ADP{} values are computed using different thresholds for each system (except \grasp{}), whereas the precision values are computed after thresholding. Due to space constraints, we excluded the experiments without results.

\section{Reevaluation}\label{app:reeval}
The results of our reevaluation over five runs using the PIDSMaker pipeline\footnote{\url{https://github.com/ubc-provenance/PIDSMaker/}} are summarized in Table~\ref{tab:reeval}. 
To increase robustness against outliers, we replaced the previous max-loss thresholding approach with a percentile-based method that uses the same calculation, substituting the maximum value with the 99.9th and 99th percentiles. 
The default hyperparameters were used to perform the test runs. 
Compared to the original evaluation, our results differ substantially. 
The results across runs are worse, and while \velox{} shows occasional detections, its FP rate remains low or negligible. 
This low FP rate further implies that behavior from unknown executables, which should be assumed to be malicious~\cite{what_we_talk}, is classified as benign.
The detection capacity increases only marginally with \velox{}, and the FP rate increases significantly.
Additionally, \orthrus{} demonstrates consistency across varying thresholding methods through its clustering, but consistently fails to detect any attacks.
The ADP metric remains constant across the thresholding variant experiments, as this is the authors' implementation~\cite{velox}, which calculates the area under the curve for multiple thresholds. 
The lack of reproducibility in the results has also been discussed by other researchers on GitHub\footnote{\url{https://github.com/ubc-provenance/PIDSMaker/issues/27}}. 
The proposed solutions involved running the code multiple times and experimenting with different hyperparameters. This does not align with our approach to reproducibility; with our code, we aim to achieve similar results across multiple runs when using the default parameters. 
Furthermore, for fairness, we conducted the experiments using the authors' specified hyperparameters. 
The experiments are marked with an H in the Table~\ref{tab:reeval}. 
Even with the optimized hyperparameters, we were unable to reproduce the results, although the results were better, and we did manage to run in some cases. 

\clearpage

\begin{table}[t]
  \centering
  \caption{Results of reevaluation and thresholding techniques for \velox{} and \orthrus{} using the \velox{} split. \label{tab:reeval}}
  \tiny
  \begin{tblr}{
    width=\linewidth,
    colspec={|X[2.5,l]|X[c]X[c]X[c]|r|X[c]X[c]X[c]|X[r]X[r]|},
    row{1,2}={valign=m,c,font=\bfseries\tiny},
    rowsep=0pt,
  }
    \hline
    \SetCell[r=2]{m} System &
    \SetCell[c=3]{c}{ADP} & & & 
    \SetCell[r=2]{m}{\ensuremath{\delta}} &
    \SetCell[c=3]{c}{Precision} & & & 
    \SetCell[c=1]{l}{TP}&
    \SetCell[c=1]{l}{FP} \\

    \SetCell[c=1]{l}{} &
    \SetCell[c=1]{c}{Mean} & 
    \SetCell[c=1]{c}{Min} & 
    \SetCell[c=1]{c}{Best} & 
    \SetCell[c=1]{l}{}&
    \SetCell[c=1]{c}{Mean} & 
    \SetCell[c=1]{c}{Min} & 
    \SetCell[c=1]{c}{Best} & 
    \SetCell[c=2]{c}{Best} \\
    \hline
    \SetRow{abovesep=1pt}\SetCell[c=10]{c}\cadetsfirstbox{}\,\cadetsEthree{} \\
    \hline
    \orthrus{}                               & \gc{0.12}0.12 &  \gc{0.09}0.09& \gc{0.14}0.14& 14\%&  \gc{0.00}0.00 & \gc{0.00}0.00& \gc{0.00}0.00 & 0 & 8   \\
    \orthrus{}\smash{\textsuperscript{\tiny 99.9th}}        & \gc{0.12}0.12 &  \gc{0.09}0.09& \gc{0.14}0.14& 14\%&  \gc{0.00}0.00 & \gc{0.00}0.00& \gc{0.00}0.00 & 0 & 8   \\
    \orthrus{}\smash{\textsuperscript{\tiny 99th}}           & \gc{0.12}0.12 &  \gc{0.09}0.09& \gc{0.14}0.14& 14\%&  \gc{0.00}0.00 & \gc{0.00}0.00& \gc{0.00}0.00 & 0 & 8   \\
    \orthrus{}\smash{\textsuperscript{\tiny H}}      & \gc{0.34}0.34 &  \gc{0.02}0.02& \gc{0.43}0.43& 24\%&  \gc{0.04}0.04 & \gc{0.00}0.00& \gc{0.06}0.06 & 1 & 17   \\
    \velox{}                                 & \gc{0.31}0.31 &  \gc{0.00}0.00& \gc{0.43}0.43& 45\%&  \gc{0.13}0.13 & \gc{0.00}0.00& \gc{0.63}0.63 & 5 & 3   \\
    \velox{}\smash{\textsuperscript{\tiny 99.9th}}           & \gc{0.31}0.31 &  \gc{0.00}0.00& \gc{0.43}0.43& 45\%&  \gc{0.01}0.01 & \gc{0.00}0.00& \gc{0.01}0.01 & 42 & 3,6k  \\
    \velox{}\smash{\textsuperscript{\tiny 99th}}             & \gc{0.31}0.31 &  \gc{0.00}0.00& \gc{0.43}0.43& 45\%&  \gc{0.00}0.00 & \gc{0.00}0.00& \gc{0.00}0.00 & 37 & 10k  \\
    \velox{}\smash{\textsuperscript{\tiny H}}   & \gc{0.36}0.36 &  \gc{0.02}0.02& \gc{1.00}1.00& 110\%&  \gc{0.18}0.18 & \gc{0.00}0.00& \gc{0.73}0.73 & 8 & 3   \\
    \hline
    \SetRow{abovesep=1pt}\SetCell[c=10]{c} \theiafirstbox{}\,\theiaEthree{} \\
    \hline
    \orthrus{}                                &\gc{0.06}0.06 &  \gc{0.04}0.04& \gc{0.08}0.08&  22\%& \gc{0.00}0.00 & \gc{0.00}0.00& \gc{0.00}0.00 & 0 & 25 \\
    \orthrus{}\smash{\textsuperscript{\tiny 99.9th}}         &\gc{0.06}0.06 &  \gc{0.04}0.04& \gc{0.08}0.08&  22\%& \gc{0.00}0.00 & \gc{0.00}0.00& \gc{0.00}0.00 & 0 & 25 \\
    \orthrus{}\smash{\textsuperscript{\tiny 99th}}           &\gc{0.06}0.06 &  \gc{0.04}0.04& \gc{0.08}0.08&  22\%& \gc{0.00}0.00 & \gc{0.00}0.00& \gc{0.00}0.00 & 0 & 25 
    \\
    \orthrus{}\smash{\textsuperscript{\tiny H}}      & \gc{0.08}0.08 &  \gc{0.02}0.02& \gc{0.15}0.15& 50\%&  \gc{0.03}0.03 & \gc{0.00}0.00& \gc{0.13}0.13 & 2 & 14   \\
    \velox{}                                  &\gc{0.32}0.32 &  \gc{0.19}0.19& \gc{0.44}0.44&  26\%& \gc{0.09}0.09 & \gc{0.07}0.07& \gc{0.10}0.10 & 12 & 106 \\
    \velox{}\smash{\textsuperscript{\tiny 99.9th}}           &\gc{0.32}0.32 &  \gc{0.19}0.19& \gc{0.44}0.44&  26\%& \gc{0.02}0.02 & \gc{0.01}0.01& \gc{0.02}0.02 & 55 & 2,0k \\
    \velox{}\smash{\textsuperscript{\tiny 99th}}             &\gc{0.32}0.32 &  \gc{0.19}0.19& \gc{0.44}0.44&  26\%& \gc{0.00}0.00 & \gc{0.00}0.00& \gc{0.00}0.00 & 59 & 12k \\
    \velox{}\smash{\textsuperscript{\tiny H}}   & \gc{0.52}0.52 &  \gc{0.03}0.03& \gc{0.90}0.90& 58\%&  \gc{0.24}0.24 & \gc{0.00}0.00& \gc{0.63}0.63 & 10 & 6   \\
    \hline
    \SetRow{abovesep=1pt}\SetCell[c=10]{c}\clearscopefirstbox{}\,\ClearscopeEthree{} \\
    \hline
    \orthrus{}                               &\gc{0.01}0.01 &  \gc{0.01}0.01& \gc{0.01}0.01&  10\%&  \gc{0.00}0.00 & \gc{0.00}0.00& \gc{0.00}0.00 & 0 & 7   \\
    \orthrus{}\smash{\textsuperscript{\tiny 99.9th}}        &\gc{0.01}0.01 &  \gc{0.01}0.01& \gc{0.01}0.01&  10\%&  \gc{0.00}0.00 & \gc{0.00}0.00& \gc{0.00}0.00 & 0 & 7   \\
    \orthrus{}\smash{\textsuperscript{\tiny 99th}}          &\gc{0.01}0.01 &  \gc{0.01}0.01& \gc{0.01}0.01&  10\%&  \gc{0.00}0.00 & \gc{0.00}0.00& \gc{0.00}0.00 & 0 & 7   \\
    \orthrus{}\smash{\textsuperscript{\tiny H}}      & \gc{0.07}0.07 &  \gc{0.01}0.01& \gc{0.11}0.11& 19\%&  \gc{0.00}0.00 & \gc{0.00}0.00& \gc{0.00}0.00 & 1 & 2030   \\
    \velox{}                                 &\gc{0.14}0.14 &  \gc{0.01}0.01& \gc{0.22}0.22&  29\%&  \gc{0.00}0.00 & \gc{0.00}0.00& \gc{0.00}0.00 & 1 & 14k   \\
    \velox{}\smash{\textsuperscript{\tiny 99.9th}}          &\gc{0.14}0.14 &  \gc{0.01}0.01& \gc{0.22}0.22&  29\%&  \gc{0.00}0.00 & \gc{0.00}0.00& \gc{0.00}0.00 & 1 & 14k   \\
    \velox{}\smash{\textsuperscript{\tiny 99th}}            &\gc{0.14}0.14 &  \gc{0.01}0.01& \gc{0.22}0.22&  29\%&  \gc{0.00}0.00 & \gc{0.00}0.00& \gc{0.00}0.00 & 1 & 14k   \\
    \velox{}\smash{\textsuperscript{\tiny H}}   & \gc{0.00}0.00 &  \gc{0.00}0.00& \gc{0.17}0.17& 146\%&  \gc{0.00}0.00 & \gc{0.00}0.00& \gc{0.00}0.00 & 1 & 14087   \\
    \hline
    \SetRow{abovesep=1pt}\SetCell[c=10]{c}        \optcfirstbox{}\,\optcfirst{} \\
    \hline
    \orthrus{}                                &\gc{0.04}0.04 &  \gc{0.02}0.02& \gc{0.06}0.06&  32\%&  \gc{0.00}0.00 & \gc{0.00}0.00& \gc{0.00}0.00 & 0 & 4   \\
    \orthrus{}\smash{\textsuperscript{\tiny 99.9th}}         &\gc{0.04}0.04 &  \gc{0.02}0.02& \gc{0.06}0.06&  32\%&  \gc{0.00}0.00 & \gc{0.00}0.00& \gc{0.00}0.00 & 0 & 4   \\
    \orthrus{}\smash{\textsuperscript{\tiny 99th}}           &\gc{0.04}0.04 &  \gc{0.02}0.02& \gc{0.06}0.06&  32\%&  \gc{0.00}0.00 & \gc{0.00}0.00& \gc{0.00}0.00 & 0 & 4   \\
    \orthrus{}\smash{\textsuperscript{\tiny H}}      & \gc{0.24}0.24 &  \gc{0.01}0.01& \gc{1.00}1.00& 157\%&  \gc{0.07}0.07 & \gc{0.00}0.00& \gc{0.33}0.33 & 1 & 2   \\
    \velox{}                                  &\gc{0.02}0.02 &  \gc{0.02}0.02& \gc{0.04}0.04&  26\%&  \gc{0.00}0.00 & \gc{0.00}0.00& \gc{0.00}0.00 & 0 & 0   \\
    \velox{}\smash{\textsuperscript{\tiny 99.9th}}           &\gc{0.02}0.02 &  \gc{0.02}0.02& \gc{0.04}0.04&  26\%& \gc{0.00}0.00 & \gc{0.00}0.00& \gc{0.00}0.00 & 11 & 2,1k \\
    \velox{}\smash{\textsuperscript{\tiny 99th}}             &\gc{0.02}0.02 &  \gc{0.02}0.02& \gc{0.04}0.04&  26\%& \gc{0.00}0.00 & \gc{0.00}0.00& \gc{0.00}0.00 & 75 & 28k  \\
    \velox{}\smash{\textsuperscript{\tiny H}}   & \gc{0.33}0.33 &  \gc{0.01}0.01& \gc{1.00}1.00& 168\%&  \gc{0.09}0.09 & \gc{0.00}0.00& \gc{0.43}0.43 & 3 & 4   \\
    \hline
    \SetRow{abovesep=1pt}\SetCell[c=10]{c} \optcsecondbox{}\,\optcsecond{}\\
    \hline
    \orthrus{}                                &\gc{1.00}1.00 &  \gc{1.00}1.00& \gc{1.00}1.00&  0\%&  \gc{1.00}1.00 & \gc{1.00}1.00& \gc{1.00}1.00 & 1 & 0   \\
    \orthrus{}\smash{\textsuperscript{\tiny 99.9th}}         &\gc{1.00}1.00 &  \gc{1.00}1.00& \gc{1.00}1.00&  0\%&  \gc{1.00}1.00 & \gc{1.00}1.00& \gc{1.00}1.00 & 1 & 0   \\
    \orthrus{}\smash{\textsuperscript{\tiny 99th}}           &\gc{1.00}1.00 &  \gc{1.00}1.00& \gc{1.00}1.00&  0\%&  \gc{1.00}1.00 & \gc{1.00}1.00& \gc{1.00}1.00 & 1 & 0   \\
    \orthrus{}\smash{\textsuperscript{\tiny H}}      & \gc{0.42}0.42 &  \gc{0.02}0.02& \gc{1.00}1.00& 146\%&  \gc{0.05}0.05 & \gc{0.00}0.00& \gc{0.25}0.25 & 1 & 3   \\
    \velox{}                                  &\gc{0.20}0.20 &  \gc{0.01}0.01& \gc{0.50}0.50&  83\%&  \gc{0.02}0.02 & \gc{0.00}0.00& \gc{0.03}0.03 & 1 & 31   \\
    \velox{}\smash{\textsuperscript{\tiny 99.9th}}           &\gc{0.20}0.20 &  \gc{0.01}0.01& \gc{0.50}0.50&  83\%&  \gc{0.01}0.01 & \gc{0.01}0.01& \gc{0.01}0.01 & 16 & 1,3k   \\
    \velox{}\smash{\textsuperscript{\tiny 99th}}             &\gc{0.20}0.20 &  \gc{0.01}0.01& \gc{0.50}0.50&  83\%&  \gc{0.02}0.02 & \gc{0.00}0.00& \gc{0.06}0.06 & 1,4k & 21k   \\
    \velox{}\smash{\textsuperscript{\tiny H}}   & \gc{0.52}0.52 &  \gc{0.03}0.03& \gc{1.00}1.00& 75\%&  \gc{0.09}0.09 & \gc{0.00}0.00& \gc{0.25}0.25 & 2 & 6   \\
    \hline
    \SetRow{abovesep=1pt}\SetCell[c=10]{c} \optcthirdbox{}\,\optcthrid{}\\
    \hline
    \orthrus{}                                &\gc{0.00}0.00 &  \gc{0.00}0.00& \gc{0.01}0.01&  49\%&  \gc{0.00}0.00 & \gc{0.00}0.00& \gc{0.00}0.00 & 0 & 15   \\
    \orthrus{}\smash{\textsuperscript{\tiny 99.9th}}         &\gc{0.00}0.00 &  \gc{0.00}0.00& \gc{0.01}0.01&  49\%&  \gc{0.00}0.00 & \gc{0.00}0.00& \gc{0.00}0.00 & 0 & 15   \\
    \orthrus{}\smash{\textsuperscript{\tiny 99th}}           &\gc{0.00}0.00 &  \gc{0.00}0.00& \gc{0.01}0.01&  49\%&  \gc{0.00}0.00 & \gc{0.00}0.00& \gc{0.00}0.00 & 0 & 15   \\
    \orthrus{}\smash{\textsuperscript{\tiny H}}      & \gc{0.01}0.01 &  \gc{0.01}0.01& \gc{0.01}0.01& 30\%&  \gc{0.00}0.00 & \gc{0.00}0.00& \gc{0.00}0.00 & 0 & 0   \\
    \velox{}                                  &\gc{0.00}0.00 &  \gc{0.00}0.17& \gc{0.50}0.50&  42\%&  \gc{0.08}0.08 & \gc{0.03}0.03& \gc{0.13}0.13 & 2 & 14  \\
    \velox{}\smash{\textsuperscript{\tiny 99.9th}}           &\gc{0.00}0.00 &  \gc{0.00}0.17& \gc{0.50}0.50&  42\%&  \gc{0.00}0.00 & \gc{0.00}0.00& \gc{0.01}0.01 & 17 & 2,9k  \\
    \velox{}\smash{\textsuperscript{\tiny 99th}}             &\gc{0.00}0.00 &  \gc{0.00}0.17& \gc{0.50}0.50&  42\%&  \gc{0.00}0.00 & \gc{0.00}0.00& \gc{0.01}0.01 & 356 & 32k  \\
    \velox{}\smash{\textsuperscript{\tiny H}}   & \gc{0.01}0.01 &  \gc{0.01}0.01& \gc{0.07}0.07& 112\%&  \gc{0.00}0.00 & \gc{0.00}0.00& \gc{0.00}0.00 & 0 & 0   \\
    \hline
  \end{tblr}

\end{table}

\section{Hyperparameter Study Results}\label{app:hyper}
Table~\ref{tab:hyper_studie} shows the results of the study. We discuss the results in Subsection~\ref{subsec:hyperparamter}.

\begin{table}[b]
  \centering
  \caption{Hyperparameter study on \cadetsEthree{} and \theiaEthree{}.\label{tab:hyper_studie}}
  \begin{tblr}{
    width=\linewidth,
    cells={font=\tiny},
    colspec={|X[2.2,l]|*{6}{X[c]}|},
    row{1}={font=\bfseries\tiny},
    hlines = {0.2pt, dashed},
    vlines = {0.2pt, dashed},
    hline{1,2,Z} = {0.5pt, solid},
    hline{3,5,6,8,9,11,12,14,15,17,18,20,21,23,24,26,27,29,30,32,33}   = {0.5pt, solid},
    vline{1,2,Z} = {0.5pt, solid},
    rowsep=0pt, 
  }
  Dataset & \meanattackrecall{}  & \meanalarms{} & \cvalarms{} & \meanh{} & \meanFone{} & \meanMacroFone{} \\
  \SetCell[c=7]{c} Defaults\\
  \cadetsfirstbox{}\,\cadetsEthree{}    &    \gc{1.00}1.00   &  \gcrREVERSE{446.8}{275.6}{575.0}446.8   &  0.33   & \gcr{21.2}{13}{23}21.2   &  \gc{0.98}0.98  & \gc{0.43}0.43 \\
  \theiafirstbox{}{}\,\theiaEthree{}    &  \gc{1.0}1.00     &  \gcrREVERSE{372.2}{254.6}{536.0}372.2   &  0.18   &  \gcr{5.0}{4}{8.0}5.0  &  \gc{0.96}0.96  & \gc{0.20}0.20\\
  \SetCell[c=7]{c} Neighborhood [20,20]\\
  \cadetsfirstbox{}\,\cadetsEthree{}    &    \gc{1.00}0.00   & \gcrREVERSE{536.0}{275.6}{575.0}536.0    &  0.13   & \gcr{18.6}{13}{23}18.6   &  \gc{0.98}0.98  & \gc{0.38}0.38\\ 
  \theiafirstbox{}{}\,\theiaEthree{}    &  \gc{1.0}1.00     &  \gcrREVERSE{377.0}{254.6}{536.0}377.0   &  0.18   &  \gcr{8.0}{4}{8.0}8.0  &  \gc{0.94}0.94  & \gc{0.20}0.20\\
  
  \SetCell[c=7]{c} Neighborhood [10000,0]\\
  \cadetsfirstbox{}\,\cadetsEthree{}    &  \gc{1.00}1.00   & \gcrREVERSE{533.0}{275.6}{575.0}533.0    &  0.06   &   \gcr{19.0}{13}{23}19.0 &  \gc{0.98}0.98  & \gc{0.45}0.45\\  
  \theiafirstbox{}{}\,\theiaEthree{}    &   \gc{1.0}1.00    &  \gcrREVERSE{448.0}{254.6}{536.0}448.0   &  0.12   & \gcr{6.6}{4}{8.0}6.6   &  \gc{0.95}0.95   & \gc{0.21}0.21\\  
  
  \SetCell[c=7]{c} Batch size 8\\
  \cadetsfirstbox{}\,\cadetsEthree{}    &   \gc{1.00}1.00   &  \gcrREVERSE{490.6}{275.6}{575.0}490.6   &   0.22  &  \gcr{16.3}{13}{23}16.3  &  \gc{0.98}0.98  & \gc{0.36}0.44\\ 
  \theiafirstbox{}{}\,\theiaEthree{}    &   \gc{1.0}1.00     &  \gcrREVERSE{330.6}{254.6}{536.0}330.6   &   0.08   &  \gcr{4.3}{4}{8.0}4.3  &  \gc{0.96}0.96  & \gc{0.20}0.20\\ 
  
  \SetCell[c=7]{c} Batch size 512\\
  \cadetsfirstbox{}\,\cadetsEthree{}    &    \gc{0.83}0.83   &   \gcrREVERSE{353.6}{275.6}{575.0}353.6  &  0.16   & \gcr{15.6}{13}{23}15.6   &  \gc{0.95}0.95  & \gc{0.22}0.22\\  
  \theiafirstbox{}{}\,\theiaEthree{}    &   \gc{1.0}1.00     &  \gcrREVERSE{254.6}{254.6}{536.0}254.6   &   0.17   &  \gcr{4.0}{4}{8.0}4.0  &  \gc{0.83}0.83  & \gc{0.07}0.07\\  
  
  \SetCell[c=7]{c} Training Epochs 2\\
  \cadetsfirstbox{}\,\cadetsEthree{}    &  \gc{1.00}1.00      &   \gcrREVERSE{448.0}{275.6}{575.0}448.0  &  0.42   &  \gcr{19.0}{13}{23}19.0  & \gc{0.98}0.98  & \gc{0.43}0.43\\
  \theiafirstbox{}{}\,\theiaEthree{}    &   \gc{1.0}1.00     &  \gcrREVERSE{303.6}{254.6}{536.0}303.6    &  0.12   &  \gcr{6.0}{4}{8.0}6.0  &  \gc{0.94}0.93  & \gc{0.20}0.20\\
  
  \SetCell[c=7]{c} Training Epochs 8\\
  \cadetsfirstbox{}\,\cadetsEthree{}    &   \gc{1.0}1.00    &  \gcrREVERSE{519.0}{275.6}{575.0}519.0   &   0.13  &  \gcr{23.0}{13}{23}23.0  &  \gc{0.98}0.98  & \gc{0.44}0.44\\
  \theiafirstbox{}{}\,\theiaEthree{}    &   \gc{1.0}1.00     &  \gcrREVERSE{351.0}{254.6}{536.0}351.0   &  0.05   &  \gcr{6.0}{4}{8.0}6.0  &  \gc{0.95}0.95  & \gc{0.20}0.20\\
  
  \SetCell[c=7]{c} Context Size 60\\
  \cadetsfirstbox{}\,\cadetsEthree{}    &    \gc{1.0}1.00   &  \gcrREVERSE{447.3}{275.6}{575.0}447.3   &  0.16   &   \gcr{15.0}{13}{23}15.0 & \gc{0.97}0.97  & \gc{0.34}0.34\\
  \theiafirstbox{}{}\,\theiaEthree{}    &    \gc{1.0}1.00    &   \gcrREVERSE{418.0}{254.6}{536.0}418.0  &  0.14   &  \gcr{5.6}{4}{8.0}5.6  &   \gc{0.94}0.94 & \gc{0.16}0.16\\ 
  
  \SetCell[c=7]{c} Context Size 15\\
  \cadetsfirstbox{}\,\cadetsEthree{}    &  \gc{1.0}1.00     &  \gcrREVERSE{275.6}{275.6}{575.0}275.6   &  0.35   &  \gcr{13.0}{13}{23}13.0  &  \gc{0.86}0.86  & \gc{0.14}0.14\\ 
  \theiafirstbox{}{}\,\theiaEthree{}    &   \gc{1.0}1.00     &  \gcrREVERSE{346.3}{254.6}{536.0}346.3   &  0.17  & \gcr{4.6}{4}{8.0}4.6   &  \gc{0.84}0.84  & \gc{0.09}0.09\\ 
  
  \SetCell[c=7]{c} Context Size 120 \& Step Size 10 \\  
  \cadetsfirstbox{}\,\cadetsEthree{}    &   \gc{1.0}1.00     &   \gcrREVERSE{479.0}{275.6}{575.0}479.0  &  0.63   & \gcr{14.0}{13}{23}14.0 &  \gc{0.98}0.98  & \gc{0.42}0.42\\ 
  \theiafirstbox{}{}\,\theiaEthree{}    &   \gc{1.0}1.00     &  \gcrREVERSE{327.3}{254.6}{536.0}327.3   &  0.15   & \gcr{4.3}{4}{8.0}4.3  &  \gc{0.95}0.95  & \gc{0.19}0.19\\ 
  
  \SetCell[c=7]{c} Context Size 60 \& Step Size 10 \\  
  \cadetsfirstbox{}\,\cadetsEthree{}    &   \gc{1.0}1.00     &   \gcrREVERSE{443.0}{275.6}{575.0}443.0  &  0.10   & \gcr{16.6}{13}{23}16.6 &  \gc{0.97}0.97  & \gc{0.33}0.33\\
  \theiafirstbox{}{}\,\theiaEthree{}    &   \gc{1.0}1.00     &  \gcrREVERSE{536.0}{254.6}{536.0}536.0   &  0.27   & \gcr{5.3}{4}{8.0}5.3  &  \gc{0.93}0.93  & \gc{0.17}0.17\\ 
  \end{tblr}
\end{table}

\end{document}